# The 2025 Roadmap to Ultrafast Dynamics: Frontiers of Theoretical and Computational Modelling

## Authors

1. **Fabio Caruso**
   Institute of Theoretical Physics and Astrophysics, University of Kiel, Leibnizstrasse 15, 24116, Kiel, Germany
   caruso@physik.uni-kiel.de

2. **Michael A. Sentef**
   Institute for Theoretical Physics and Bremen Center for Computational Materials Science, University of Bremen, 28359 Bremen, Germany
   Max Planck Institute for the Structure and Dynamics of Matter, Center for Free-Electron Laser Science (CFEL), Luruper Chaussee 149, 22761 Hamburg, Germany
   sentef@uni-bremen.de

3. **Claudio Attaccalite**
   CNRS/Aix-Marseille Université, Centre Interdisciplinaire de Nanoscience de Marseille UMR 7325 Campus de Luminy, 13288 Marseille cedex 9, France.
   claudio.attaccalite@univ-amu.fr

4. **Michael Bonitz**
   Institute of Theoretical Physics and Astrophysics, University of Kiel, Leibnizstrasse 15, 24116, Kiel, Germany
   bonitz@theo-physik.uni-kiel.de

5. **Claudia Draxl**,
   Physics Department and CSMB Adlershof, Humboldt-Universität zu Berlin, 12489 Berlin, Germany
   claudia.draxl@physik.hu-berlin.de

6. **Umberto De Giovannini**
   Università degli Studi di Palermo, Dipartimento di Fisica e Chimica—Emilio Segrè, Palermo I-90123, Italy
   umberto.degiovannini@unipa.it

7. **Martin Eckstein**
   Institute for Theoretical Physics, University of Hamburg, Notkestraße 9-11, 22607 Hamburg, Germany and The Hamburg Centre for Ultrafast Imaging, Hamburg, Germany
   martin.eckstein@uni-hamburg.de

8. **Ralph Ernstorfer**
   Fritz Haber Institute of the Max Planck Society, Faradayweg 4-6, 14195 Berlin, Germany.
   Institut für Optik und Atomare Physik, Technische Universität Berlin, Straße des 17 Juni 135, 10623 Berlin, Germany.
   ernstorfer@tu-berlin.de

9. **Michael Fechner**
   Max Planck Institute for the Structure and Dynamics of Matter, Hamburg, Germany
   michael.fechner@mpsd.mpg.de

10. **Myrta Grüning**
    School of Mathematics and Physics, Queen's University Belfast (BT7 1NN University Road, Belfast Northern Ireland).
    m.gruening@qub.ac.uk

11. **Hannes Hübener**
    Max Planck Institute for the Structure and Dynamics of Matter and Center for Free Electron Laser Science, 22761 Hamburg, Germany
    hannes.huebener@mpsd.mpg.de

12. **Jan-Philip Joost**
    Institute of Theoretical Physics and Astrophysics, University of Kiel, Leibnizstrasse 15, 24116, Kiel, Germany
    joost@theo-physik.uni-kiel.de

13. **Dominik M. Juraschek,**
    Department of Applied Physics and Science Education, Eindhoven University of Technology, Eindhoven, Netherlands
    d.m.juraschek@tue.nl

14. **Christoph Karrasch**
    Technische Universität Braunschweig, Institut für Mathematische Physik, Mendelssohnstr. 3, 38106 Braunschweig, Germany
    c.karrasch@tu-braunschweig.de

15. **Dante Marvin Kennes**
    Institut für Theorie der Statistischen Physik, RWTH Aachen University and JARA – Fundamentals of Future Information Technology, 52056 Aachen, Germany
    Max Planck Institute for the Structure and Dynamics of Matter, Center for Free Electron Laser Science, 22761 Hamburg, Germany
    dante.kennes@mpsd.mpg.de

16. **Simone Latini**
    Department of Physics, Technical University of Denmark, 2800 Kgs. Lyngby, Denmark
    Max Planck Institute for the Structure and Dynamics of Matter and Center for Free-Electron Laser Science, Luruper Chaussee 149, Hamburg 22761, Germany
    simola@dtu.dk

17. **I-Te Lu**
    Max Planck Institute for the Structure and Dynamics of Matter and Center for Free-Electron Laser Science, Luruper Chaussee 149, Hamburg 22761, Germany
    i-te.lu@mpsd.mpg.de

18. **Ofer Neufeld**
    Schulich Faculty of Chemistry, Technion - Israel Institute of Technology, 320000 Haifa, Israel
    ofern@technion.ac.il

19. **Enrico Perfetto**
    University of Rome Tor Vergata, via della Ricerca Scientifica 1, 00133 Rome (Italy)
    enrico.perfetto@roma2.infn.it

20. **Laurenz Rettig**
    Fritz Haber Institute of the Max Planck Society, Faradayweg 4-6, 14195 Berlin, Germany.
    rettig@fhi-berlin.mpg.de

21. **Ronaldo Rodrigues Pela,**
    Supercomputing Department, Zuse Institute Berlin (ZIB), Berlin, Takustraße 7, 14195 Berlin,

Germany
ronaldo.rodrigues@zib.de

22. **Angel Rubio**
    Max Planck Institute for the Structure and Dynamics of Matter and Center for Free Electron Laser Science, 22761 Hamburg, Germany
    Center for Computational Quantum Physics (CCQ), The Flatiron Institute, 162 Fifth avenue, New York NY 10010.
    angel.rubio@mpsd.mpg.de

23. **Joseph F. Rudzinski,**
    Physics Department and CSMB Adlershof, Humboldt-Universität zu Berlin, 12489 Berlin, Germany
    joseph.rudzinski@physik.hu-berlin.de

24. **Michael Ruggenthaler**
    Max Planck Institute for the Structure and Dynamics of Matter and Center for Free-Electron Laser Science, Luruper Chaussee 149, Hamburg 22761, Germany
    michael.ruggenthaler@mpsd.mpg.de

25. **Davide Sangalli**
    Istituto di Struttura della Materia (ISM - CNR), Area della Ricerca di Roma 1, Monterotondo Scalo, Italy.
    davide.sangalli@ism.cnr.it

26. **Michael Schüler**
    PSI Center for Scientific Computing, Theory and Data, 5232 Villigen PSI, Switzerland and Department of Physics, University of Fribourg, CH-1700 Fribourg, Switzerland.
    michael.schueler@psi.ch

27. **Samuel Shallcross**
    Max-Born-Institut für Nichtlineare Optik und Kurzzeitspektroskopie, 12489 Berlin, Germany
    Samuel.Shallcross@mbi-berlin.de

28. **Sangeeta Sharma**
    Max-Born-Institut für Nichtlineare Optik und Kurzzeitspektroskopie, 12489 Berlin, Germany
    Institute for theoretical solid-state physics, Freie Universität Berlin, Arnimallee 14, 14195 Berlin, Germany
    Sangeeta.Sharma@mbi-berlin.de

29. **Gianluca Stefanucci**
    University of Rome Tor Vergata, via della Ricerca Scientifica 1, 00133 Rome (Italy)
    gianluca.stefanucci@roma2.infn.it

30. **Philipp Werner**
    Department of Physics, University of Fribourg, 1700 Fribourg, Switzerland
    philipp.werner@unifr.ch

## Abstract

The exploration of ultrafast phenomena is a frontier of condensed matter research, where the interplay of theory, computation, and experiment is unveiling new opportunities for understanding and engineering quantum materials. With the advent of advanced experimental techniques and computational tools, it has become possible to probe and manipulate nonequilibrium processes at unprecedented temporal and spatial resolutions, providing insights into the dynamical behavior of matter under extreme conditions. These capabilities have the potential to revolutionize fields ranging from optoelectronics and quantum information to catalysis and energy storage.

This Roadmap captures the collective progress and vision of leading researchers, addressing challenges and opportunities across key areas of ultrafast science. Contributions in this Roadmap span the development of ab initio methods for time-resolved spectroscopy, the dynamics of driven correlated systems, the engineering of materials in optical cavities, and the adoption of FAIR principles for data sharing and analysis. Together, these efforts highlight the interdisciplinary nature of ultrafast research and its reliance on cutting-edge methodologies, including quantum electrodynamical density-functional theory, correlated electronic structure methods, nonequilibrium Green's function approaches, quantum and ab initio simulations.

# Content

# 1 – Introduction

**Fabio Caruso,** Institute of Theoretical Physics and Astrophysics, University of Kiel, Leibnizstrasse 15, 24116, Kiel, Germany
caruso@physik.uni-kiel.de

**Michael A. Sentef,** Institute for Theoretical Physics and Bremen Center for Computational Materials Science, University of Bremen, 28359 Bremen, Germany. Max Planck Institute for the Structure and Dynamics of Matter, Center for Free-Electron Laser Science (CFEL), Luruper Chaussee 149, 22761 Hamburg, Germany
sentef@uni-bremen.de

The investigation of ultrafast phenomena in laser-driven quantum materials has become a frontier topic in modern condensed matter research, offering unparalleled possibilities to flexibly probe and control the properties of materials on extreme time and length scales. Advances in experimental techniques, such as time-resolved spectroscopies and ultrafast imaging, combined with breakthroughs in computational modeling and data-driven approaches, are enabling researchers to drive, observe, and understand nonequilibrium phases ranging from spin, charge and orbital ordering phenomena via topological phases and putative excitonic condensates to possible driven superconductivity.

Over the past decade, the field has developed through a refinement of experimental techniques that now allow far more precise and targeted excitation protocols than ever before. These advancements open new venues towards properties on demand [Bas17] enabled by light-matter coupling. Critically, several of the most interesting phenomena in driven materials occur in situations where already at equilibrium the structural and electronic properties are challenging to model theoretically. The underlying reason is that they are governed by the quantum-many body interplay of correlation [Roh11], topology [Sie19], many-body interactions, or quantum nuclear effects [Nov19] [Li19]. This interplay offers rich opportunities for the emergence of novel classes of ultrafast phenomena or optical control. Yet, their complexity obviously becomes an even bigger challenge when out-of-equilibrium conditions are established, possibly resulting in the emergence of light-induced phases, hybrid light-matter states [Bao22], or transient topological behavior [McI20].

This challenge, in turn, has spurred theoretical and computational developments aimed at providing microscopic insights and guiding principles for nonthermal pathways to ultrafast control in quantum materials [Tor21]. Recent advances in the modeling of ultrafast phenomena have bridged critical gaps in the fundamental theory of light-matter coupling and many-body interactions, while also contributing to the development of first-principles codes, computational workflows, and FAIR (findable, accessible, interoperable, reusable) data protocols for nonequilibrium phenomena [Sch22]. The diversity and complementarity of these activities makes it opportune to take stock of the current status of the field of modeling ultrafast phenomena in quantum materials.

The treatment of electric fields within a quantum-electrodynamical framework has revealed new prospects to model hybrid light-matter states resulting, e.g., from cavity embedding [Hub20][Gar21][Sch22] [Blo22] and Floquet dressing [Oka19] [Rud20]. Significant progress has been made in treating many-body interactions and correlations in driven systems, via nonequilibrium generalizations of dynamical mean-field theory [Aok14], diagrammatic methods for electron-exciton-phonon-spin coupling [Ste23], and neural quantum states [Car17]. These advances provide new opportunities for the development of theoretical spectroscopy frameworks, capable of predictively simulating the outcome of complex time-dependent measurements. New directions have emerged to explore nonlinear structural [Man16] and optical responses in strong fields as routes for engineering materials properties or optically induce structural and topological phase transitions. The development of standardized frameworks and metadata schemes are facilitating the interoperability and reuse of complex datasets in time-resolved spectroscopies and simulations. Figure 1 highlights some of these activities.

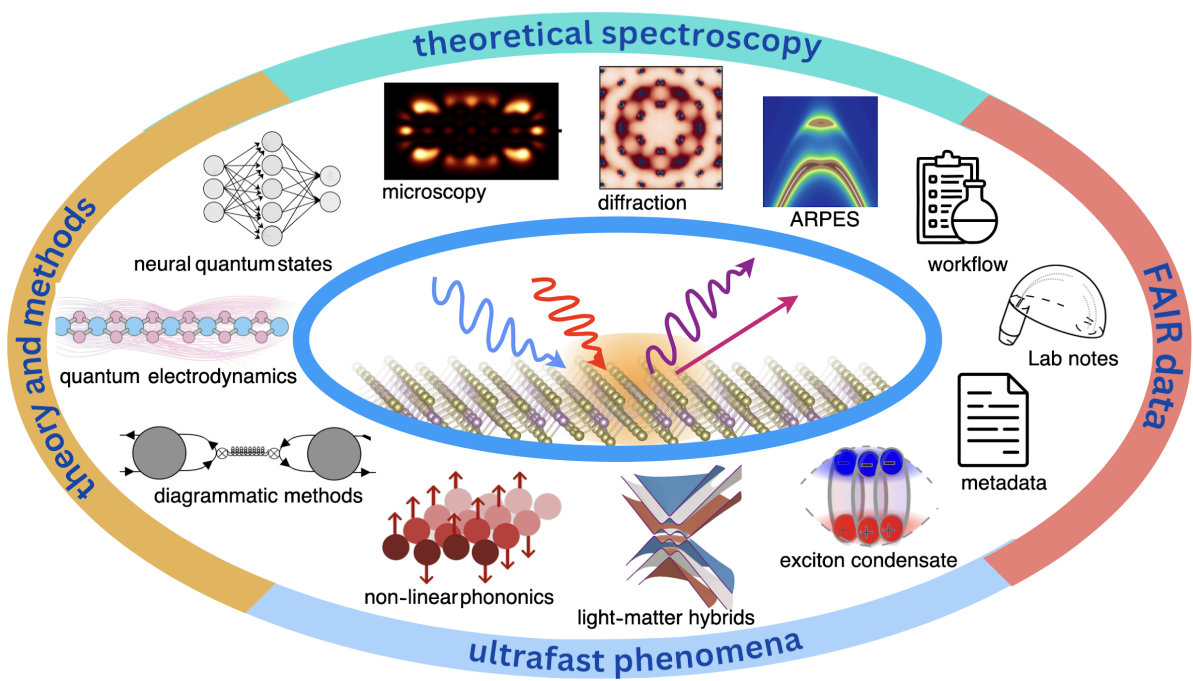

Overall, the field of modeling ultrafast phenomena in quantum materials is at an exciting juncture, marked by rapid advancements in theoretical frameworks, computational techniques, and data-driven approaches. The integration of novel methods, such as quantum-electrodynamical treatments, neural quantum states, and nonequilibrium many-body theories, is unveiling unprecedented potential for understanding and controlling light-matter interactions on ultrafast timescales. However, significant challenges remain, particularly in bridging theory and experiment, modeling complex driven systems, and establishing robust frameworks for handling time-resolved data. This Roadmap aims to offer an overview of the diverse and complementary research efforts driving advancements in the theoretical and computational modeling of ultrafast phenomena. These include the development of many-body Green's function approaches for studying ultrafast phenomena (sections 2–5), new strategies for tailoring material properties through quantum electrodynamics and Floquet engineering (sections 6–8), ab initio techniques for investigating exciton, phonon, and magnetization dynamics (sections 9–12), and the establishment of new protocols and infrastructures to implement FAIR principles for managing time-resolved materials data (sections 13–14). Contributors are experts across different areas, spanning the developments of theory and methods, ultrafast phenomena, theoretical spectroscopy frameworks, and FAIR data protocols. Besides highlighting recent advances in the field, this joint effort further aims to identify open problems and key challenges whose solution will further advance the field.

## Acknowledgements

*F.C. acknowledges funding by the Deutsche Forschungsgemeinschaft (DFG), projects number 443988403 and 499426961. M.A.S. was funded by the European Union (ERC, CAVMAT, project no. 101124492).*

## 2 – Nonequilibrium excitons from first-principles many-body theory

**Enrico Perfetto**, University of Rome Tor Vergata, via della Ricerca Scientifica 1, 00133 Rome (Italy)
enrico.perfetto@roma2.infn.it

**Gianluca Stefanucci,** University of Rome Tor Vergata, via della Ricerca Scientifica 1, 00133 Rome (Italy)
gianluca.stefanucci@roma2.infn.it

*Keywords: nonequilibrium Green functions, diagrammatic methods, electron-phonon interactions*

**Status**

Transition-metal dichalcogenides, layered heterostructures, and other two-dimensional materials are highly promising for next-generation optoelectronic devices [1,2]. The weak dielectric screening in these materials promotes the formation and coexistence of well-defined, optically bright zero-momentum singlet excitons [3,4], as well as dark inter-valley and triplet excitons [5-7]. A deeper understanding of the fundamental mechanisms behind exciton formation and scattering is essential for guiding research and accelerating advancements in this field.

Laser pulses with subgap frequencies excite bright excitons, transferring the coherence of the light to the excitonic states. The dynamics of the driven system is well described by the so-called Hartree plus screened exchange approximation (HSEX), which is essentially a mean-field theory where the Fock potential is calculated with a statically screened Coulomb interaction [8]. During and shortly after the optical excitation, the nonequilibrium state consists of coherent (or virtual) excitons [9] and possibly coherent optical phonons [10]. In fact, for pulse durations of a few tens of femtoseconds or shorter, the force acting on the nuclei is displacive, maximizing the amplitude of the phonon modes coupled to the excitons. In this coherent regime, conduction states are dressed by coherent excitons, and leave distinctive fingerprints in time-resolved (TR) ARPES. Subgap excitonic sidebands consisting of valence bands hybridized with conduction bands appear [11-13]. For low excitation densities these are replicas of the valence bands, blue-shifted by the exciton energy, with an intensity proportional to the exciton wavefunction. With increasing the excitation densities the replicas flatten and eventually acquire the shape of a mexican hat. The coherent regime persists up to the timescale of the excitonic lifetimes [15], which typically shorten with increasing temperature and excitation density. Coherent excitons can be detected using four-wave-mixing experiments [5,14]. An alternative approach, yet to be realized experimentally, involves conducting TR-ARPES experiments using ultrashort probes. In this scenario, while the sidebands may fade, the signal intensity oscillates as a function of the pump-probe delay, with frequencies corresponding to the exciton energies [12].

A first-principles many-body theory for the coherent-to-incoherent crossover and for the incoherent regime, applicable even beyond the weak pumping scenario, is under development. Current approaches rely on model Hamiltonians that incorporate pre-screened electron-electron and electron-phonon couplings, denoted as *v* and *g*. In reconciling these model-based descriptions with fundamental ab initio results—such as the Bethe-Salpeter equation—an overscreening of *v* and *g* emerges. Nonetheless, these approaches provide important insights on the exciton dynamics. The charge imbalance created by the laser pulse sets the nuclear lattice in motion. In addition to exciting coherent optical phonons, this process also changes the populations of the phonon modes. Electron-phonon scattering is responsible for destroying exciton coherence [16,17] and diffusing excitons [18–19]. Bright coherent excitons are then converted into bright and dark incoherent excitons. The finite excitation density is responsible for screening the electron-hole attraction and shrinking the gap, thereby renormalizing the exciton energies. This effect is well captured in TR optical experiments. The concomitant photobleaching of the exciton peaks also provides valuable information, as its magnitude grows with the population of excitons. For too large excitation densities the exciton peaks are quenched, a typical signature of the excitonic Mott transition.

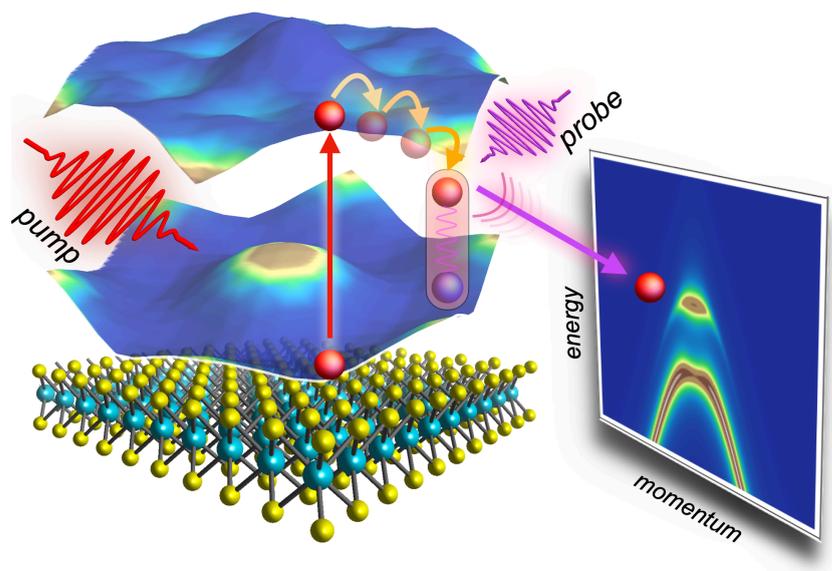

**Figure 1.** (Left) Schematic illustration of dynamical creation of excitons after photo-excitation by a pump pulse in a monolayer transition metal dichalcogenide. (Right) ARPES spectrum generated by a delayed probe pulse revealing the excitonic replica of the valence band.

Moreover optical experiments are sensitive to the presence of coherent optical phonons. In fact, coherent oscillations of the nuclear lattice along the normal modes survive for much longer than coherent excitons, as the dominant decoherence mechanism is phonon-phonon scattering. These oscillations cause a time-dependent modulation of the exciton peaks, which has been observed [22,23] and theoretically explained [10]. Optical experiments are not suited to investigate dark excitons nor to resolve excitonic states in momentum space. In contrast, TR-ARPES is sensitive to bright, momentum-dark and spin-dark excitons. Furthermore, the TR-ARPES signal does not distinguish between coherent or incoherent excitons. Therefore, similarly to the previously described scenario of the coherent regime, in the incoherent regime a small population of excitons with momentum $Q$ and energy $E_{XQ}$, gives rise to a replica of the valence bands which are shifted in both energy and momentum by $E_{XQ}$ and $Q$ respectively [24,25]. Furthermore, the signal intensity is related to the wavefunction of both bright and dark excitons.

**Current and Future Challenges**

Using the first-principles Hamiltonian for electrons and phonons [26] as a starting point, we have recently derived the excitonic Bloch equations (XBE) for the coupled dynamics of coherent and incoherent excitons [27]. The XBE are free from overscreening issues and capture both the initial transient regime driven by external laser fields and the transition from the coherent to the incoherent regime. The XBE provide a minimal set of equations to describe exciton formation, diffusion, and thermalization at moderate excitation densities. The XBE description can, and in some cases must, be improved along several directions that we outline below: (i) A first challenge is to account for the time-dependent renormalization of the exciton energies and screened interaction due to the finite excitation density. Closely related to this is the need to go beyond the Tamm-Dancoff approximation; (ii) Another important challenge is the overcoming of the Markovian approximation through schemes like the time-linear formulation of the Kadanoff-Baym equations (see also section 3) [28,29]. Within these frameworks it is also possible to relax the ubiquitous static screening for both the electron-electron and electron-phonon interactions [30,31]; (iii) The interaction between excitons and coherent optical phonons presents a third key challenge for the future [10]. It has been pointed out that it involves *bare* electron-phonon couplings. Extracting bare couplings from ab initio codes using pseudopotentials poses conceptual problems; (iv) A less critical challenge consists in abandoning the frozen phonon approximations along the lines outlined in Ref.

[32,33]. However, the phononic self-energy must be refined to account for excitonic effects in the polarization; (v) At large enough excitation density the Coulomb mediated exciton-exciton scattering can no longer be ignored. This interaction has been investigated using model Hamiltonians [34,35], but a first principles description is currently lacking; (vi) The same applies to anharmonic effects like the phonon-phonon and Debye-Waller interactions; (vii) Non-diagonal exciton-exciton correlations have been shown to be critical for a proper description of the excitonic lifetimes [36]. Their inclusion in a real-time framework has not yet been explored; (viii) Developing a unifying framework to simulate the real-time dynamics of semiconductors and insulators that is applicable even for above-gap photoexcitations and incorporates all the aforementioned improvements, represents the most significant challenge. The main difficulty lies in a consistent treatment of free carriers (electrons and holes) and excitons as the latter are composite carriers.

**Concluding Remarks**

The coupled exciton-phonon dynamics involves processes such as screening buildup, carrier dressing by phonons, phonon dressing by carriers, transient lattice heating, exciton formation and diffusion, and intra- and inter-valley scattering. Many of these processes are effective already during the coherent-to-incoherent crossover and cannot be treated as independent. A first-principles real-time description of the relevant degrees of freedom still has a long way to go. Crucial concepts and methods have already been developed, and others are certainly yet to come.

Steering excitons with electric, magnetic, or optical fields is a central topic in modern science due to its profound implications for a wide range of technologies, including optoelectronics, photovoltaics, and photocatalysis. Research in this area offers the possibility for enhancing the performance of energy-harvesting devices, light-emitting materials, and next-generation quantum devices. Beyond its technological relevance, this research is also crucial for advancing our understanding of quantum matter in nonequilibrium conditions. In fact, the dynamics of nonequilibrium excitons and phonons opens new avenues for exploring the fundamental principles of quantum coherence, dissipation, and phase transitions in many-body systems.

**Acknowledgements**

This work has been supported by MIUR PRIN (Grant No. 2022WZ8LME), INFN through the TIME2QUEST project, Tor Vergata University through Project TESLA and the European Union's Horizon Europe research and innovation programme under the Marie Sklodowska-Curie grant agreement 101118915 (project TIMES).

## 3 – New algorithms and linear scaling approaches to NEGF simulations

**Jan-Philip Joost,** Institut für Theoretische Physik und Astrophysik, Christian-Albrechts-Universität zu Kiel, 24098 Kiel, Germany Kiel Nano, Surface and Interface Science KiNSIS, Kiel University, Germany
joost@physik.uni-kiel.de,

**Michael Bonitz,** Institut für Theoretische Physik und Astrophysik, Christian-Albrechts-Universität zu Kiel, 24098 Kiel, Germany Kiel Nano, Surface and Interface Science KiNSIS, Kiel University, Germany
bonitz@physik.uni-kiel.de

*Keywords: nonequilibrium Green functions, diagrammatic methods, nonequilibrium methods*

**Status**

Electronic correlations play a central role in 2D quantum materials, in particular, for geometrically confined finite clusters, such as graphene nanoribbons (GNR), where the density of states (DOS) becomes space dependent. An adequate theoretical description requires nonequilibrium Green functions (NEGF) simulations, see, e.g., Refs. [1-3].

Figure 1 compares differential conductance measurements at the GNR edge and bulk to DOS computations within different approximations. While uncorrelated models miss important spectral features, our NEGF-*GW* results show excellent agreement with the experiment [4]. When such structures are exposed to external excitation, rapid dynamics on (sub-)femtosecond time scales are induced. An example is the impact of highly charged ions that creates a strongly non-uniform field in the material leading to charge redistribution, electron transfer to the ion and very strong secondary electron emission (SEE) into vacuum [5]. Interestingly, measurements of SEE revealed strong differences (factor 6) between graphene and $MoS_2$, cf. left part of Fig. 2. Again, NEGF simulations provide good agreement (right figure part) and, in addition, an explanation of the physical origins for this difference: the higher mobility of electrons in graphene and the larger lattice constant and electronic correlations in $MoS_2$. Moreover, NEGF are perfectly suited to fully account for electronic correlation effects under arbitrary nonequilibrium conditions allowing to compute all properties, including the spectral function and density of states, with (sub-)femtosecond time and sub-nm space resolution.

However, NEGF simulations are computationally very costly: the CPU time effort scales cubically with the simulation duration $N_t$. This can be improved to $N^2_t$ – scaling, by restricting the time propagation of the two-time Green functions to the time diagonal, using the Hartree-Fock-Generalized Kadanoff-Baym ansatz (HF-GKBA) [3,6,7], however, only for the simplest correlation selfenergy – the second Born approximation (SOA). Very recently we developed the G1-G2-scheme that exactly reformulates the HF-GKBA in time-local form. This cuts the CPU time to linear scaling in $N_t$ [8] and has since changed the field of NEGF simulations. The G1-G2 scheme makes very long simulations possible and achieves linear scaling also for *GW* and *T*-matrix approximations [9]. It even allowed us to selfconsistently combine, for nonequilibrium conditions, *GW* and particle-particle and particle-hole *T*-matrix diagrams into the dynamically screened ladder approximation (DSL) [10]. Further, it gives direct access to 2-particle observables, such as pair distributions, see Ref. [11], for a recent overview. The G1-G2-scheme has already been applied to a variety of systems (see also section 2) [12-14].

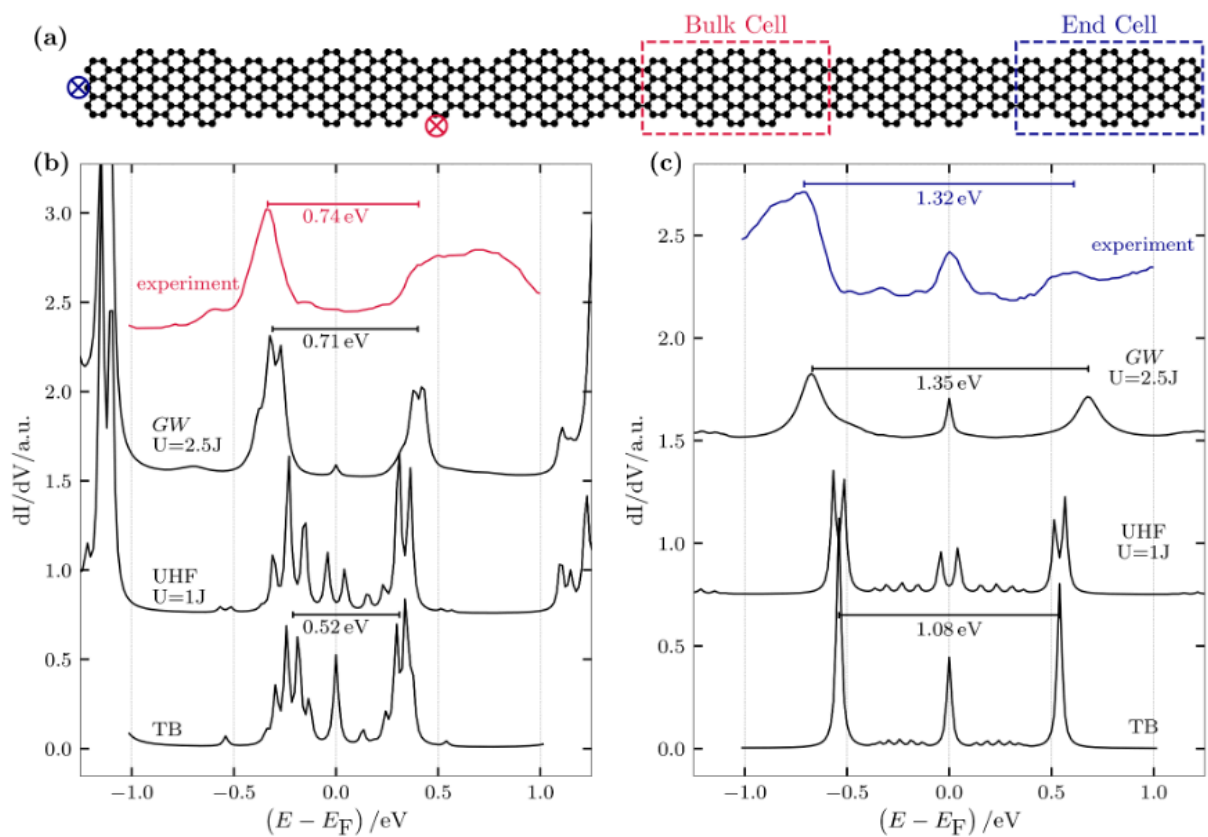

**Figure 1.** Measured differential conductance (colored line) for two positions (bulk and end) in the GNR heterostructure shown on top. Comparison to simulations using tight binding (TB), unrestricted Hartree-Fock (UHF), end NEGF simulations with GW selfenergy. The figure is reproduced from Ref. [4] with permission from ACS.

**Current and Future Challenges**

Aside from the CPU-time advantage, the G1-G2 scheme which couples time-local equations for the one-particle (G1) and correlated part of the two-particle NEGF (G2) also has disadvantages and critical issues that have to be solved. The first group of problems is associated with the observation that under certain circumstances the G1-G2 scheme can lead to increased computational effort:

1. As a two-particle quantity G2 is a rank-4 tensor. This leads to increased memory consumption for large systems, compared to standard NEGF, and a, in general, worse scaling with the basis dimension [15].
2. Calculations may become unstable (especially for the DSL selfenergy), at strong coupling, and may require additional regularization procedures to guarantee stability ("purification" [10]) which further worsens the basis scaling.
3. Applications of the G1-G2 scheme so far concentrated on lattice models. First tests for uniform systems (k-space basis) indicate an unfavorable scaling of G2 with the basis dimension [11,16]. Applications to other basis sets, such as atomic orbitals or Kohn-Sham states, are straightforward but have yet to be explored.

The second group of more general issues is inherited by the G1-G2 scheme from the HF-GKBA:

1. The linear scaling approach (as the standard HF-GKBA) loses access to high-quality correlated spectral functions which is a major advantage of the NEGF approach [15].

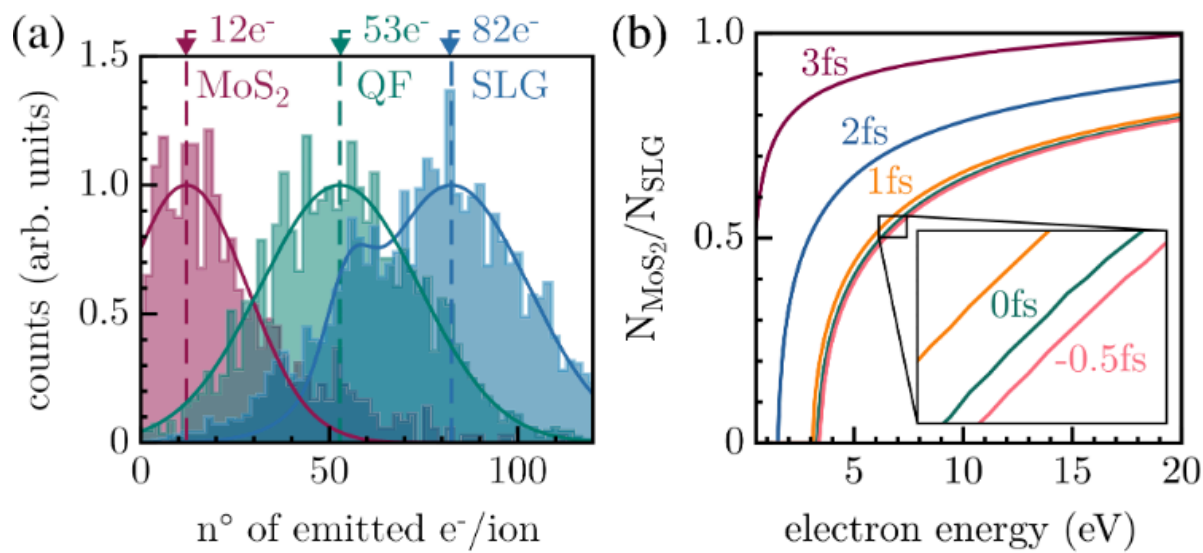

**Figure 2.** Secondary electron emission from monolayers of $MoS_2$ and graphene upon impact or a highly charged ion. Left: measured distribution of emitted electrons. Right: time evolution of the ratio of emitted electrons from NEGF-simulations (time zero refers to ion penetration of the monolayer). Figure reproduced from Ref. [5].

2. Correlated ground states cannot be generated via the imaginary branch (Matsubara component) but have to be generated by adiabatically switching on the interaction, which may cause a significant CPU time overhead [11].

**Advances in Science and Technology to Meet Challenges**

There are several promising ways in which the challenges listed above could be addressed. The numbering of the following points does not necessarily coincide with the one in the previous section. For mitigating the issue of unfavorable basis scaling the following concepts have already been proposed and tested, in part:

1. The dimensionality of G2 can be reduced using embedding concepts where only a reduced set of states or sites is simulated with full accuracy (i.e., with a high level selfenergy) and the rest ("the environment") is treated simpler, e.g., on the HF level where G2 equals zero. The embedding idea has already been integrated into the G1-G2 scheme, and it was demonstrated that time linear scaling can be preserved [17]. At the same time correlation effects have not yet been treated fully selfconsistently.
2. Within the NEGF community there is currently extensive work being done on the topic of "tensor decomposition" and "low-rank compression" (see also section 5) [18,19]. Applying this idea to G2 could help to reduce the basis scaling.
3. The most promising solution seems to be the recently developed quantum fluctuations approach (QFA) [20-22] which can be seen as a physically (instead of mathematically) motivated way of compressing G2. In this approach the G1-G2 scheme is reformulated in terms of fluctuations of field operator products and, in many cases, G2 can be eliminated (expressed in terms of one-particle fluctuations). First tests of the QFA with *GW* selfenergies for lattice models have been very promising, as the simulations were found to be stable and, thus, do not suffer from the increased scaling due to purification.

Likewise, there are several ideas under active development to address the more general issues:

1. A major task is to restore spectral information in the G1-G2 scheme. One way is to use Koopmans' theorem or the extended Koopmans' theorem and extend it to nonequilibrium [15]. Another promising route is to include a time-off-diagonal propagation into the scheme to deduce the spectrum from the time propagation [23]. Here various preliminary concepts exist that need to be tested and implemented.
2. More generally, there exist various machine learning concepts that could be invoked to either derive improved selfenergy approximations or to extend simulations to larger systems [15].

**Concluding Remarks**

To summarize, the development of the linear scaling time-local G1-G2-scheme [8-11] has changed the situation of NEGF simulations for correlated electron systems – a development that has only started. The scheme has the capability to combine the power of the NEGF approach (that is traditionally extremely CPU-time consuming) with computational efficiency that matches that of competing approaches such as time-dependent DFT. Not surprisingly, this dramatic acceleration does not come for free: the price to pay is to cope with large memory requirements for storing G2, in particular, when it comes to dealing with physically relevant systems. On the other hand, the quantum fluctuations approach [20-22] which is currently under active development is expected to solve the majority of the problems. The enormous reduction of the numerical cost observed by the QFA, without loss of accuracy, is expected to make it applicable to a vast range of systems and nonequilibrium problems, including laser-excited (sub)femtosecond electron dynamics in quantum materials, with (sub-)nanometer space resolution.

**Acknowledgements**

We acknowledge discussions with the members of our group, in particular Erik Schroedter and Christopher Makait. This work is supported by DFG via project BO1366-16.

# 4 – Nonequilibrium dynamical-mean field theory and beyond, driven strongly correlated systems with local and nonlocal correlations

**Martin Eckstein,** *Institute for Theoretical Physics, University of Hamburg, Notkestraße 9-11, 22607 Hamburg, Germany and The Hamburg Centre for Ultrafast Imaging, Hamburg, Germany*
martin.eckstein@uni-hamburg.de

**Philipp Werner,** *Department of Physics, University of Fribourg, 1700 Fribourg, Switzerland*
philipp.werner@unifr.ch

*Keywords: nonequilibrium Green functions, dynamical mean field theory, impurity solvers*

**Status**

Dynamical mean field theory (DMFT) provides an approximate description of lattice systems which becomes exact in the noninteracting, atomic and infinite-dimensional limits [Georges1996]. By focusing on the effects of local correlations, DMFT enables a mapping of the lattice system onto a self-consistently defined effective single-site model (an impurity problem), which makes numerical calculations tractable. DMFT captures the Mott transition and has been widely used for equilibrium studies of correlated lattice models and (in combination with density functional theory) for the simulation of strongly correlated materials [Kotliar2006]. The DMFT formalism can be generalized to nonequilibrium setups by solving the DMFT equations on the Keldysh contour [Aoki2014] using nonequilibrium Green's functions (NEGF). Nonequilibrium DMFT simulations have been used to study ultrafast processes in correlated lattice systems [Aoki2014], such as photo-induced insulator-metal transitions, carrier localization in strong fields, dynamical phase transitions and nonthermal fixed points [Tsuji2013, Marino2022]. More recently, DMFT simulations of quasi-steady states have been used to address various types of long-lived nonthermal electronic orders [Murakami2023]. The significant effect of nonlocal correlations on the photo-carrier dynamics was demonstrated with cluster extensions of DMFT [Eckstein2016], while the role of nonlocal charge fluctuations and dynamical screening was clarified by combining DMFT with the nonlocal GW diagrams (GW+DMFT formalism) [Golez2017]. NEGF simulations based on a local approximation have also provided insights into the effects of electron-phonon couplings on the electronic relaxation after photo excitation [Sentef2013].

Most existing nonequilibrium studies addressed the basic physics of simple model systems, such as the single-band Hubbard model, and only a few recent works started to tackle the complexities of material-specific predictions. With the available technology, a handshake between theory and experiment is often difficult to achieve. Simulations are typically limited to short times below 1 ps, only a few orbitals can be treated, and in contrast to equilibrium, we lack numerically exact nonequilibrium impurity solvers. Most previous nonequilibrium DMFT studies were based on weak- or strong-coupling perturbative solvers. Still, these simulations have been helpful for the interpretation of experiments on photo-excited Mott insulators [Ligges2019], and a benchmark against a cold atom simulator revealed an almost quantitative agreement [Sandholzer2019]. With further technical improvements on the solver side, diagrammatic extensions of DMFT, and the embedding of DMFT or GW+DMFT into ab initio frameworks, nonequilibrium DMFT-based methods should reach a level of realism and accuracy which enables direct comparisons to experiments on correlated materials.

**Current and Future Challenges**

*Multi-scale behavior in time*: A key challenge in nonequilibrium DMFT simulations of correlated lattice systems and materials is describing dynamics on vastly different timescales. The excitation process in laser driven lattice systems can occur on the timescale of the inverse hopping. During multi-cycle pulses, the system may reach a Floquet state with properties distinct from those of the equilibrium system. After the pulse, relaxation processes can lead to long-lived prethermalized states due to approximate symmetry constraints, or relaxation bottlenecks which result from a restricted scattering phase space [Murakami2023]. Long-range orders potentially bring in additional timescales through the dynamics of collective modes and the proximity to nonthermal fixed points. The full thermalization process thus often occurs on timescales which are orders of magnitude longer than the excitation process. Bridging these timescales in a single simulation is a major numerical challenge. A description of this multi-scale dynamics in time is essential to understand even conceptually the physical mechanisms behind metastable photo-induced orders.

*Nonlocal correlations and spatial inhomogeneities*: Cluster and diagrammatic extensions of nonequilibrium DMFT in principle allow one to capture nonlocal correlations. However, the computational cost for these schemes is significantly higher, so that with current technology, only small clusters [Eckstein2016] and simple diagrammatic extensions such as GW+DMFT [Golez2017] have been implemented. Methods involving four-point correlation functions require a large amount of memory and may be applicable only to nonequilibrium steady-states. Inhomogeneous systems with surfaces, interfaces or multi-layer structures can be efficiently treated with nonequilibrium real-space DMFT [Eckstein2013], although the memory demand for such calculations becomes substantial.

*Material-specific calculations*: Nonequilibrium DMFT calculations using model parameters derived from ab initio simulations have been realized, but this approach misses the nontrivial interplay between correlation and screening effects. This feedback can in principle be handled by the GW+DMFT method, which has been successfully applied to model systems [Golez2017]. A full ab initio embedding however leads to dynamically screened interactions which vary over an energy range of several tens of eV, which means that very high energy processes need to be retained in the time evolution. Material realistic simulations within DMFT also create a need for more sophisticated impurity solvers, which can handle multiple sites or orbitals, and retarded density-density or spin-spin interactions. Most of the existing nonequilibrium DMFT applications are based on simple solvers such as weak-coupling perturbation theory or the non-crossing approximation [Aoki2014].

**Advances in Science and Technology to Meet Challenges**

*Long-time simulations*: A promising approach to addressing the challenge of multi-(time)-scale simulations using Kadanoff-Baym equations emerges from the application of advanced data compression techniques to nonequilibrium Green's functions (NEGF). These include hierarchical decomposition schemes for two-time matrices, tensor-train representations following a logarithmic time discretization [Shinaoka2023], and high-order exponential fits utilizing linear prediction and matrix pencil algorithms. The overarching challenge lies in combining memory-efficient data representations with an effective simulation protocol. Successfully overcoming these challenges could also enable accurate quantum kinetic simulations based on realistic band structures, such as real-time GW simulations.

*Impurity solvers*: Real-time Monte Carlo methods [Cohen2015] provide numerically exact impurity solvers for some applications. Beyond that, the development of impurity solvers for DMFT could benefit significantly from recent advances in tensor network (TN) techniques. TNs approximate the dependence of functions on many variables, with tensor cross-interpolation [Fernandez2022] offering a method to construct compressed TN representations by sampling functions at suitably chosen pivot points. This approach could facilitate systematic diagrammatic calculations, which otherwise face the challenge of high dimensionality in both temporal integrals and diagram topologies. Moreover, neural-network-inspired data structures and optimization algorithms developed in the artifical intelligence community hold potential for further efficiency gains.

*Nonequilibrium phase transitions*: In the context of light-driven phase transitions, the interplay between lattice dynamics and electronic structure will likely play a crucial role in stabilizing non-thermal states of matter. On the DMFT side, semiclassical simulations of electron-lattice dynamics have been explored in simplified settings, including global lattice displacements [Varma2024] and the emergence of dynamically generated disorder [Picano2023]. Progress in understanding photo-induced phase transitions could stem from integrating nonequilibrium DMFT with advances in molecular dynamics simulations [Schobert2024], which facilitate efficient modelling of dynamics in weakly correlated charge density wave systems.

*Link to experiments*: On the experimental side, significant progress has been made in probing the fastest timescales, extending into the attosecond regime. For correlated electrons, unique nonlinear dynamics which is distinct from the nearly collision-less short-time behavior in good metals can be anticipated on these timescales [Valmispilt2024]. This dynamics may have implications for light-driven electronics. Moreover, it offers a new perspective for a quantitative connection between nonequilibrium DMFT and experiments, as atomic positions remain effectively frozen on these timescales. A second critical experimental development involves advancements in time-resolved X-ray spectroscopies [Mitrano2024], which can open up novel avenues for probing the local electronic structure accessible through DMFT simulations.

## Concluding Remarks

In summary, the initial development of nonequilibrium DMFT has demonstrated its potential to uncover complex ultrafast phenomena from photo-induced phase transitions to dynamical phase transitions. However, significant challenges persist. Tackling multi-scale temporal dynamics, incorporating nonlocal correlations, and enhancing material-specific accuracy demand substantial advancements in methodology. An interesting perspective both for nonequilibrium DMFT impurity solvers and for NEGF simulations in general is provided by tensor-train representations of real-time Green's functions. The capabilities of DMFT to provide quantitative, material-specific insights will deepen particularly through the real-time GW+DMFT formalism, and through the integration of DMFT with semiclassical lattice dynamics simulations. Coupled with experimental progress, e.g., in time-resolved X-ray and sub-fs spectroscopy, these simulations may help to unlock new nonequilibrium phenomena in correlated materials and pave the way towards light-driven control of quantum materials.

## Acknowledgements

M.E. acknowledges funding through the Cluster of Excellence "CUI: Advanced Imaging of Matter" of the DFG – EXC 2056 – project ID 390715994. P.W. acknowledges support by SNSF Grant No. 200021-196966.

## 5 – Driven strongly correlated systems in one and two dimensions

**Christoph Karrasch,** Technische Universität Braunschweig, Institut für Mathematische Physik, Mendelssohnstr. 3,
38106 Braunschweig, Germany
c.karrasch@tu-braunschweig.de

**Dante Marvin Kennes,** Institut für Theorie der Statistischen Physik, RWTH Aachen University and JARA – Fundamentals of Future Information Technology, 52056 Aachen, Germany
Max Planck Institute for the Structure and Dynamics of Matter, Center for Free Electron Laser Science, 22761 Hamburg, Germany
dante.kennes@mpsd.mpg.de

*Keywords: nonequilibrium many-body physics, low-dimensional quantum systems, driven quantum matter*

**Status**

The last decades have seen tremendous developments in laser technologies [1] which, among many other important applications [2], allow one to address the inherent degrees of freedom in solids in a near surgical way. This opens up ultrafast control pathways to coherently excite specific electronic, phononic, or other collective degrees of freedom, which is in marked contrast to simply using lasers as a source of (incoherent) local heating. First examples for ultrafast coherent control of basic, effectively noninteracting properties of solids include those of electronic band structure [3] or lattice dynamics [4], including effective modification of the average lattice constant. This provides novel control routes complementary to equilibrium means such as altering chemistry (e.g., to affect the band structure) or external pressure (e.g., to change the lattice constant). However, arguably the most tantalizing aspects of solid state physics relate to the fact that "more is different" [5] and that emergent phenomena take center stage in condensed matter research. Ultrafast control of emergent phenomena such as topology, superconductivity, or quantum magnetism opens up possibilities to disentangle their intricate nature, deepening our fundamental understanding, and also to implement technologically relevant functionalities [6].

However, describing quantum matter in the presence of strong correlations that trigger emergent phenomena is a daunting task already in equilibrium. Questions about how these phenomena are modified or possibly even induced by driving thus poses a formidable challenge, which often requires one to address the intricate interplay of correlated quantum particles on various spatial and temporal scales. Particularly, in one- and two-dimensional quantum systems the relative strength of quantum and thermal fluctuations lead to a delicate balance of emergent ordering tendencies and fluctuation-induced disordering. Many attempts were undertaken to develop methods that can deal with such strongly-correlated nonequilibrium problems in the past, each with its specific strengths and limitations: Exact diagonalization is restricted to small system sizes. Tensor networks can treat large one-dimensional systems but are limited by the amount of entanglement and hence restricted to short time scales [7]. Methods based on Keldysh-Schwinger Green's functions [8] are less limited in terms of dimensionality but require an approximation for the two-time self-energy associated with interactions. Low-order perturbation theory [9] becomes challenging with higher orders, and in particular in low dimensions their applicability is often limited by infrared divergences that hail the onset of emergent phenomena at low temperatures. Renormalization group approaches such as the functional renormalization group (FRG) can systematically improve on this last problem but also suffer from quickly exacerbated complexity if one goes beyond leading orders. Alternative approximation strategies are subject of active investigation see [...] and section 3. Dynamical mean field theory (DMFT) approximates the self-energy by a spatially-local version [10, 11], which, however, tends to be particularly problematic in the low-dimensional limit. This can be partly remedied by out-of-equilibrium DMFT cluster extensions (see section 4), but the usefulness in one and two dimensions is still unclear. More generally, powerful equilibrium methods do not always extend to nonequilibrium straightforwardly. Quantum Monte Carlo samplings of nonequilibrium quantum many-body problems are very often plagued by a dynamical version of the infamous sign problem [12]. Numerical renormalization group techniques, which are limited to zero-dimensional systems, can be extended to nonequilibrium [13] but numerical implementations are complicated by the lack of an energy-scale-separation guiding principle. Novel methods based on machine learning techniques are opening up a fascinating perspective [14], but have yet to establish their full potential and delineate its limitations in the field.

To summarize, the field of "ultrafast phenomena in condensed matter" is currently at the intriguing crossroads of quickly-maturing experimental technologies and urgently-required improvements to theoretical methods. The latter is key to moving away from intuition-driven setups that currently prevail to theory-guided ones. In this short perspective, we highlight some examples for high-impact methods-driven developments with a particular focus on one- and two-dimensional systems.

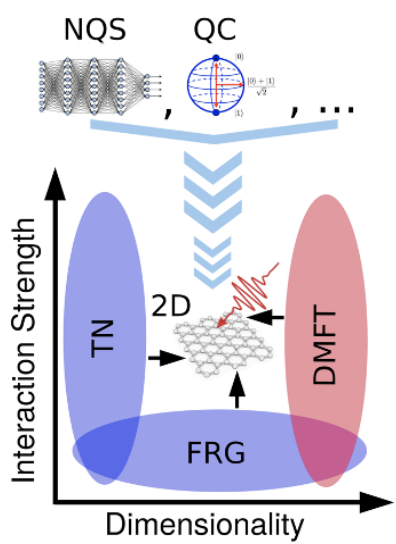

**Figure 1:** The faithful computation of the dynamics of low-dimensional quantum systems relies on a multi-method approach. Tensor network (TN), functional renormalization group (FRG), and dynamical mean field theory (DMFT) are already used synergistically to approach the challenging two-dimensional regime from different angles for which these complementary methods can be applied. With new emerging computational approaches such as, e.g., those relying on neural quantum states (NQS), or quantum computing (QC), novel complementary computational inroads are currently under intense investigation, and the quantum dynamics of two-dimensional systems provides a prime application.

## Current and Future Challenges

There are a plethora of challenges in the description of driven one- and two-dimensional correlated quantum systems, in fact, too many to outline them here. Instead we focus on two:

*Challenge One: A Myriad of Timescales*

Emergent solid state phenomena feature a range of relevant timescales. The time scale of the "bare" degrees of freedom (e.g., electronic or phononic ones) is often much faster than the one associated with collective quasiparticles behavior or with emergent ordering tendencies (e.g., superconductivity), which typically reside at much smaller energies (and are thus slower). Nonequilibrium setups are even more complex when emergent phenomena such as amplitude and phase dynamics of orders adhere to fundamentally different timescales. This is particularly true in low dimensions where strong effects of fluctuations can easily dislodge such orders. Any faithful theoretical approach thus needs to be capable of dealing with timescales inherent to the driving pulse (usually very fast) all the way to the emergent dynamical timescales (usually relatively slow). Insights from both transient dynamics and the long-time steady drive are just as integral as understanding how the one transitions to the other [15].

*Challenge Two: The Complexity of Heating*

In order to faithfully describe the ultrafast dynamics in quantum many-body systems (such as solids), one needs to capture inelastic scattering during the drive, which will deposit energy (heat) into the system. It is vital to understand the delicate balance between the coherent engineering of emergent phenomena and the counteracting detrimental incoherent heating. However, many simple techniques such as the Hartree-Fock approximation do not include inelastic processes and are therefore inadequate for describing the ultrafast dynamics of emergent phenomena in solids on timescales longer than the typical scales associated with inelastic scattering. Including heating in diagrammatic expansions requires a next-to-leading order description, while in tensor networks one strays away from low entanglement – both are challenging.

| Method | Strengths | Challenges |
|---|---|---|
| Exact Diagonalization | exact dynamics | exponential scaling in system size |
| Tensor Networks [7] | exact dynamics | 2D, dynamical entanglement build-up |
| Green Func.: Perturbative [9] | flexible w.r.t dim., transient & steady | accurate description of heating |
| Green Func.: DMFT [10, 11] | non-perturbative, transient & steady | non-local correlations, impurity solver |
| Quantum Monte Carlo [12] | exact dynamics | dynamical sign problem |
| Numerical RG [13] | exact dynamics | beyond 0d, hierarchy energy/time scales |
| Neural Quantum States [14] | exact dynamics, flexibility | practicalities currently unknown |
| … | | |

**Table 1:** Rough characterization of computational methods. Focusing on the description of driven strongly correlated systems in one and two dimensions, we tabulate some methodological strengths and limitations of the approaches discussed.

## Advances in Science and Technology to Meet Challenges

We will focus on two important developments that are required to address the challenges outlined above, pertaining to two separate lines of thought on how to improve our theoretical description of ultrafast phenomena in one- and two-dimensional quantum systems.

*Extensions to Diagrammatic Expansions and a Case Study of FRG*

Methods based on Keldysh Green functions are generally not restricted by dimensionality and can thus tackle one and two dimensional systems alike. More importantly, such approaches can be set up in the real time domain to describe transient dynamics, in the steady state, or for periodically driven Floquet systems. They hence provide a natural way to study multiple time scales (Challenge One). The key quantity is the self-energy associated with two-body interactions U, which is often computed using (Feynman) diagrammatic techniques in a way that is perturbative in U. In order to describe heating (Challenge Two), one needs to go beyond leading-order schemes, which is demanding.

One particular diagrammatic technique is the functional renormalization group, which implements Wilson's general RG idea by reformulates a given many-body problem analytically in terms of an infinite hierarchy of flow equations for vertex functions (such as the self-energy) with an infrared cutoff serving as the flow parameter [16]. This hierarchy is truncated perturbatively in U. Depending on the context, the ensuing schemes can be viewed as a 'RG enhanced Hartree-Fock approach', as a 'self-consistent random-phase approximation', or as a 'resummation of leading logs'. Leading-order FRG schemes have been used to tackle the nonequilibrium steady state [17, 18], the real time dynamics [19], and Floquet problems [20]. The steady-state has been studied via second-order schemes which incorporate heating; e.g., it was shown how nonequilibrium dynamics and dissipation can conspire to control the phase diagram of one-dimensional quantum systems [21]. Generalizing this approach to the real-time domain or to Floquet problems is a subject of current research. Open quantum systems can be addressed straightforwardly at no additional cost, and problems with bosons (e.g., a coupling to phonons) pose no fundamental hurdle (see also sections 2 and 3).

*Simulating the Wavefunction: Of Matrix Product States, Neural Quantum States and Digital Quantum Computation*

Tensor networks (TNs) provide an efficient way to encode a wave function whose bi-partite entanglement in some local basis is small. In one dimension, rigorous theorems guarantee small entanglement in the ground state of certain systems, which can then be computed variationally using the famous density-matrix renormalization group algorithm [7]. The real-time evolution of a given initial wavefunction (e.g., the ground state of a different Hamiltonian, a simple product state, or one related to correlation functions) can be simulated very accurately using a variety of different TN-based algorithms such as the time-dependent variational principle or the time-dependent block decimation (see [22] for a comprehensive review). The key bottleneck is the build-up of entanglement in the time-evolved state, which often limits the simulation to short time scales. In practice, only one-dimensional or thin two-dimensional systems (e.g., few-leg ladders) are accessible using these algorithms; TN schemes can also be set up directly in two-dimensions but have similar limitations [23]. TNs can be used to tackle open systems (Challenge Two) in a Lindblad formulation, and a variety of approaches to tackle bosonic infinite-dimensional local Hilbert spaces exist [24].

The entanglement bottleneck of traditional tensor networks has stimulated a variety of novel research avenues. Neural quantum states were introduced recently [14, 25, 26]; certain random neural network states feature volume law entanglement [27] and might thus be a fruitful tool to simulate the dynamics. First results are promising, but a deeper understanding of the virtues and shortcomings of this method is lacking. Combining the insights gleaned from neural networks and relating them to, e.g., tensor network approaches will hopefully fill this gap. Another current avenue of research is to rephrase neural quantum states directly in terms of tensor network structures, which holds the promise to push current limitations [28] and to enter uncharted realms of quantum many-body dynamics.

Simulating the dynamics of a generic quantum many body system is a hard problem, and complexity-theoretic arguments suggest that it might be out of reach of any classical simulation [29]. Following Richard Feynman's famous advice "if you want to make a simulation of nature, you'd better make it quantum mechanical", an alternative ansatz is to directly employ a quantum simulator. E.g., in the field of ultracold gases, first examples of such (so-called analog) simulations were already performed successfully [30]. In contrast, digital quantum simulation, i.e., a simulator that can be used for any Hamiltonian $H(t)$, is still in its relative infancy. In the noisy intermediate-scale quantum era [31], digital quantum simulation can currently only shed light on specific cases such as those involving excessive noise. However, recent advances in error correction [32] raise hope of achieving fault tolerance in digital quantum simulation, and the physics of driven one- and in particular two-dimensional quantum systems seem a landmark future application.

## Concluding Remarks

The future of ultrafast phenomena in condensed matter is bright, in particular in the context of the tantalizing emergent physics of low-dimensional quantum systems. With experimental advances quickly spearheading into uncharted realms, this offers a unique opportunity for theory to provide a framework that guides the field in the future. Key methodological advances are crucial, which will integrally combine insights and methodologies from traditional, successful techniques such as those building on quantum many-body Green's functions with those of novel emerging ansätze approaching at the horizon of the modern computational condensed matter landscape, such as neural quantum states and analog and digital quantum simulation. Driven strongly correlated systems in one and two dimensions provide an ideal arena for these efforts that take place at the interface of past and future methodological endeavors. This will not only deepen our understanding of the fundamental physics at play but will also open up future technological opportunities, e.g., in the context of utilizing low dimensional materials in on-chip optoelectronic devices.

## 6 – First-principles Floquet analysis of materials engineering

**Hannes Hübener,** Max Planck Institute for the Structure and Dynamics of Matter and Center for Free Electron Laser Science, 22761 Hamburg, Germany
hannes.huebener@mpsd.mpg.de

**Ofer Neufeld,** Schulich Faculty of Chemistry, Technion - Israel Institute of Technology, 320000, Haifa, Israel
ofern@technion.ac.il

**Umberto De Giovannini,** Università degli Studi di Palermo, Dipartimento di Fisica e Chimica—Emilio Segrè, Palermo I-90123, Italy
umberto.degiovannini@unipa.it

**Angel Rubio,** Max Planck Institute for the Structure and Dynamics of Matter and Center for Free Electron Laser Science, 22761 Hamburg, Germany
Center for Computational Quantum Physics (CCQ), The Flatiron Institute, 162 Fifth avenue, New York NY 10010.
angel.rubio@mpsd.mpg.de

*Keywords: ultrafast dynamics, ultrafast spectroscopy, Floquet engineering*

**Status**

First-principles Floquet analysis (1) is an approach to study spectroscopic features of solid-state systems subjected to time-periodic perturbations, such as those induced by intense laser fields. Floquet theory states that under such conditions the time evolution of the driven system can be described by a quasi-static Hamiltonian that exhibits eigenvalues (that appear as sidebands) and a modified spectrum with respect to the groundstate. Importantly, Floquet eigenstates can have fundamentally different properties than the groundstate, and hence analyzing strongly driven systems in terms of Floquet theory can provide a suitable framework to understand their non-equilibrium properties. By combining Floquet theory with *ab initio* computational methods – most notably time-dependent density functional theory (TDDFT) and many-body perturbation theory – one can gain insight into the electronic structures and excitation dynamics of materials, revealing the origins of novel spectroscopic features.

In recent years, significant progress has been made in applying first-principles Floquet analysis. Notably, the method has been instrumental in predicting dynamical band engineering, where the electronic band structure of a material is altered on an ultrafast timescale by external lasers. The success of Floquet analysis is based on the synergy between theoretical developments and experimental techniques, especially time-resolved angle-resolved photoemission spectroscopy (TR-ARPES), which has been instrumental in validating the predictions of first-principles Floquet calculations (2; 3; 35) and predicting novel phenomena such as the presence of electron-phonon couplings (4). TR-ARPES allows for the direct observation of transient electronic states and band structures with attosecond temporal resolution, providing empirical evidence for phenomena like Floquet sidebands and dynamical band inversions predicted by theory (5). Pioneering ARPES experiments showed that Floquet states in solids are not only detectable, but also lead to a transiently modified bandstructure (6; 7) in topological insulators and more recently Floquet analysis has been instrumental in expanding this concept to insulators (8). Floquet analysis can also explain the modification of nonlinear optical properties in pump-probe spectroscopies (9; 10).

The primary focus of this field lies in the exploration of light-induced topological phases and the control of electronic properties via Floquet engineering. Theoretical studies have demonstrated that by modulating the amplitude, frequency, and polarization of the driving field, one can manipulate the topological invariants of a system, leading to phases with exotic properties like anomalous Hall effects and protected edge modes (11; 12) These predictions open up possibilities for developing optoelectronic devices that can be dynamically tuned post-fabrication and be employed in petahertz electronics (13). There has been major progress in observing Floquet topology experimentally (14; 10). However, it remains challenging due to the difficulty of isolating unique signatures of Floquet states compared to complex nonequilibrium effects, e.g., due to requirements of dynamical Floquet population control, especially in graphene, which has been the workhorse of theory predictions.

First-principles Floquet analysis has been extended to study the nonequilibrium electronic structure not only under laser driving but also under other coherent excitations such as phonons (15; 16). Notably, coherent excitons are predicted with Floquet analysis to lead to a sizable renormalisation of the electronic bandstructure in materials with strong excitonic effects (17; 18; 19).

Recent efforts have also focused on integrating Floquet band engineering with machine learning (20) and optimization (21) techniques to engineer nonequilibrium phases with determined Floquet properties. These approaches can pave the way to realizing novel Floquet states of matter that are otherwise difficult to engineer, e.g., flat bands or negative effective masses.

**Current and Future Challenges**

Several challenges impede the full realization of first-principles Floquet analysis in practical applications. Computational cost is one of the main factors that prevents first-principle Floquet analysis from being widely applied. Modeling realistic materials under periodic driving requires solving the many-body time-dependent Schrödinger equation, which is computationally intensive, especially for correlated systems. While methods like TDDFT provide a framework for including these interactions, they often rely on approximations that may not capture all relevant phenomena (e.g., the employed level of electronic interactions or the adiabatic approximation in TDDFT). This limitation can lead to discrepancies between theoretical predictions and experimental results, particularly in materials where excitonic effects and electron-electron interactions play a significant role (18; 17). However, inherently including the full Brillouin zone and multiple bands can have very important implications often missed by simple models that focus on a specific k-space region or feature of the system(22), which necessitates overcoming these challenges.

The nonequilibrium nature of driven systems introduces additional complexity in modeling dissipation and decoherence mechanisms. Interactions with phonons, impurities, and other environmental factors that introduce decoherence and hence counteract the Floquet phases can significantly alter the system's response. Methods to account for the complex time-resolved nonequilibrium many-body electronic structure under such conditions have been developed based on nonequilibrium Green's functions (23; 24). However, using these methods as a fully first-principles framework remains a major challenge in the field, especially if long time scales are required to account for the coherent nature of the Floquet dressing process as well as dephasing and scattering (14).

One key issue in first-principles Floquet analysis is the lack of a well-defined approach for determining electronic occupations within the Floquet framework (1), which complicates predictions and often necessitates approximations that are sensitive to the specifics of the pump field. Additionally, the method generates a large number of Floquet sidebands which are redundant beyond a fixed energy window. This increases computational complexity and obscures analysis. Efficient and reliable techniques are essential to reduce the redundancy, yet implementing such downfolding or the solution of the Floquet eigenvalue problem in a restricted energy range remains challenging within a first-principles context, limiting the precision and applicability of current Floquet-based predictions.

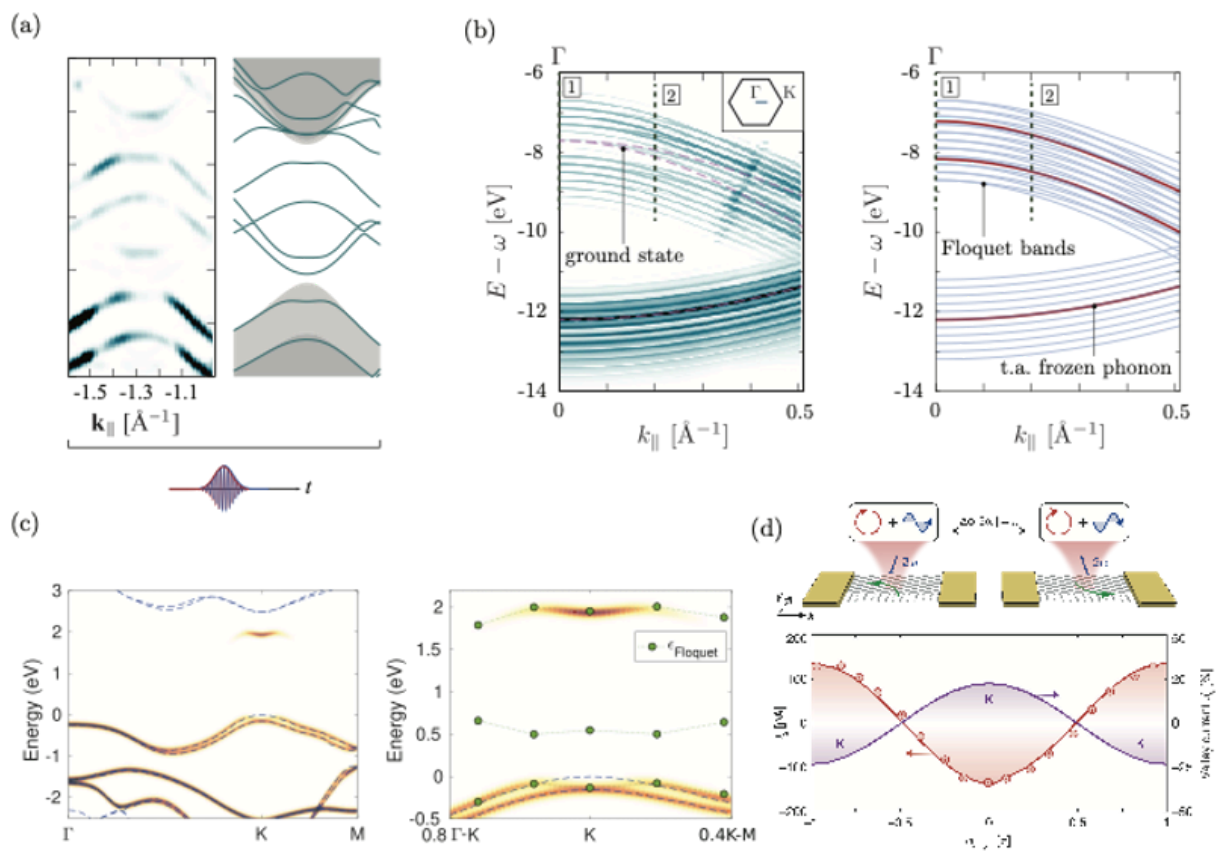

**Figure 1.** Examples of first principles Floquet analysis in the literature. (a) Simulated time-resolved ARPES spectrum of WSe$_2$ with overlapping pump and probe pulses (left panel) show photoemission features in the gap region that correspond to Floquet sidebands (right panel). Adapted from (2) (b) Simulated time-resolved ARPES spectrum of Graphene with a coherently excited phonon mode. Due to the strong electron-phonon coupling multiplessidebands are seen (left panel) to be forming around the equilibrium bands (dashed lines). Such features cannot be explained by frozen phonon band calculations (right panel, red) but are instead a perfect match for the sidebands calculated from Floquet analysis (blue lines). Adapted from (15) (c) Simulated ARPES spectrum of MoS$_2$ with a coherently excited exciton. The large exciton binding energy not only leads to the appearance of a shadow band feature in the gap region (left panel) but also strongly renormalizes the top valence band in agreement with Floquet analysis (right panel) adapted from (17) (d)adapted from (10).

An often underappreciated challenge is the accurate description of the electromagnetic environment the material creates and in which the material resides (25). Traditional models employing the Fresnel equations are limited to interfaces between homogeneous media and may not adequately capture the complexities of nanostructured or heterogeneous environments. In the context of Floquet analysis it becomes essential to go beyond the Fresnel framework, integrating Maxwell's equations with first-principles electronic structure methods to account for phenomena such as near-field effects, plasmonic resonances, and the influence of substrates or surrounding media on the material's optical properties. Failure to accurately model the electromagnetic environment can lead to significant errors in predicting spectroscopic signatures and device performance.

Another frontier in the field is the shift from merely analyzing Floquet properties to actively engineering and realizing them. Designing materials and devices that exploit Floquet-induced phenomena requires precise control over the driving field parameters—such as frequency, amplitude, and polarization—and an in-depth understanding of how these parameters influence the system's electronic states. A possible way towards this control is to combine nonequilibrium driving of materials with cavities, where photonic density of states is engineered to facilitate a desired frequency distribution of the driven phase, enabling tunable material phenomena (see section 7). This endeavor is challenging because it involves a complex interplay between material properties and external fields, and also since the cavity acts as an instigator for multiple photonic channels that need to be numerically described. Moreover, experimental realization of such engineered systems demands advances in fabrication techniques to create structures that can withstand intense fields and maintain coherence over the required timescales. Notably, approaches using tailored light and multiple driving laser frequencies have recently had substantial success, and are also expected to emerge as prominent Floquet sources (26; 27; 28).

**Advances in Science and Technology to Meet Challenges**

Addressing the challenges in first-principles Floquet analysis requires efforts in both theoretical and numerical developments. Significant advances are being made to overcome computational limitations, improve the modeling of many-body interactions and electromagnetic environments, and shift from the analysis to the engineering of Floquet properties.

Besides the more obvious technological improvements such as in parallel computing and graphics processing units there are emerging computational paradigms that will help to meet the challenges facing first principles Floquet analysis. Time-dependent density matrix renormalization group (t-DMRG) methods and tensor network approaches are being explored to reduce the computational complexity of solving the time-dependent Schrödinger equation (see section 5). Additionally, machine learning algorithms are being integrated to create surrogate models that approximate system behaviors, thereby reducing the need for exhaustive calculations without compromising accuracy (29). Above all the advent of utility in the realm of quantum computing will certainly provide a new level of versatility to the method (30).

Advancements in theoretical methods will enhance the accurate accounting of many-body interactions and electron correlation effects. Improved exchange-correlation functionals and phenomenological description of decoherence based on the Lindblad equation in time-dependent density functional theory (TDDFT) are being developed to better capture electron-electron and electron-hole interactions. Furthermore, nonequilibrium Green's function methods are improving their computational complexity (31) (see section 3) and will in the future provide a robust first-principles framework for modeling electron correlations in driven systems, enabling more precise predictions of excitonic effects and other many-body phenomena.

Accurately describing the electromagnetic environment requires going beyond traditional Fresnel equations. Integrating Maxwell's equations with first-principles electronic structure methods is a significant advancement in this area. Coupled TDDFT-Maxwell techniques (32) allow for a simultaneous description of the electromagnetic field and the time-dependent electronic structure capturing near-field effects, plasmonic resonances, and substrate influences. Notably, this will also allow describing quantum-electrodynamical processes that occur under ultrastrong light-matter coupling conditions (25), opening a new field of hybrid light-matter effects in Floquet materials (33; 34).

**Concluding Remarks**

First-principles Floquet analysis stands at the forefront of exploring, controlling, and harnessing the dynamic properties of solids under nonequilibrium conditions. While challenges remain in computational demands and theoretical limitations, ongoing advancements are steadily overcoming these hurdles. The synergy between improved computational methods, theoretical innovations, and experimental validation is key to unlocking new states of matter and their functionalities. As the field progresses, we can anticipate the emergence of materials and devices that leverage these dynamic excitations for applications in electronics, photonics, and quantum technologies. Continued investment in research and collaboration across disciplines will be essential in realizing the full potential of Floquet engineering in solid-state and photonic systems.

## Acknowledgements

*We acknowledge support from European Union – Next Generation EU through project THENCE – Partenariato Esteso NQSTI – PE00000023 – Spoke 2, and from the Marie Sklodowska-Curie Doctoral Networks TIMES grant No. 101118915 and SPARKLE grant No. 101169225. The Flatiron Institute is a division of the Simons Foundation.*

## 7 – Cavity materials engineering and QEDFT

**I-Te Lu,** Max Planck Institute for the Structure and Dynamics of Matter and Center for Free-Electron Laser Science, Luruper Chaussee 149, Hamburg 22761, Germany
i-te.lu@mpsd.mpg.de

**Simone Latini,** Department of Physics, Technical University of Denmark, 2800 Kgs. Lyngby, Denmark
Max Planck Institute for the Structure and Dynamics of Matter and Center for Free-Electron Laser Science, Luruper Chaussee 149, Hamburg 22761, Germany
simola@dtu.dk

**Michael Ruggenthaler,** Max Planck Institute for the Structure and Dynamics of Matter and Center for Free-Electron Laser Science, Luruper Chaussee 149, Hamburg 22761, Germany
michael.ruggenthaler@mpsd.mpg.de

**Angel Rubio,** Max Planck Institute for the Structure and Dynamics of Matter and Center for Free-Electron Laser Science, Luruper Chaussee 149, Hamburg 22761, Germany
Center for Computational Quantum Physics (CCQ), The Flatiron Institute, 162 Fifth avenue, New York, NY 10010, USA
angel.rubio@mpsd.mpg.de

**Status**

Cavity materials engineering, distinct from polaritonic physics or driven quantum matter, uses photon-field fluctuations within a cavity to modify the *ground state* of embedded solid-state materials in the absence of real photons. This approach builds on demonstrations that photon-field fluctuations, predicted by quantum electrodynamics (QED), can alter molecular properties and chemical reactions. These demonstrations are at the core of the emerging field of polaritonic chemistry. While theoretical methods exist for polaritonic chemistry, the extension to solid-state materials is a recent advancement.

Theoretical studies suggest that in idealized cavities, quantum-vacuum fluctuations of photons can induce ferroelectric and ferromagnetic phase transitions, control topological properties, and manipulate superconductivity [1], [2], [3], [4], [5]. Although scarce at such an early stage, pioneering experiments on solid-state materials have enabled the manipulation of properties at equilibrium, such as the quantum Hall effect [6], [7], metal-to-insulator charge-density-wave transitions in 1T-TaS$_2$ [8], and Dicke physics in condensed matter [9]. These results have been enabled by advances in design of cavities with adjustable strong light-matter coupling and photon mode frequencies.

These findings suggest a broader approach for controlling materials using light, without relying on potentially damaging lasers. By introducing a novel 'light-matter' dimension with a dark cavity, we can alter material properties in equilibrium through photon-field fluctuations. This dimension complements traditional control parameters, such as pressure, temperature, and doping, offering a new pathway to achieve desirable material properties.

To bridge gaps between experimental measurements and theoretical predictions for realistic photon-coupled materials, quantum-electrodynamical density-functional theory (QEDFT), offers certain advantages over other methods. It effectively manages complex, realistic materials by handling many electronic states, treats electrons and photons equally, and avoids the exponential scaling problem that increases computational costs [1].

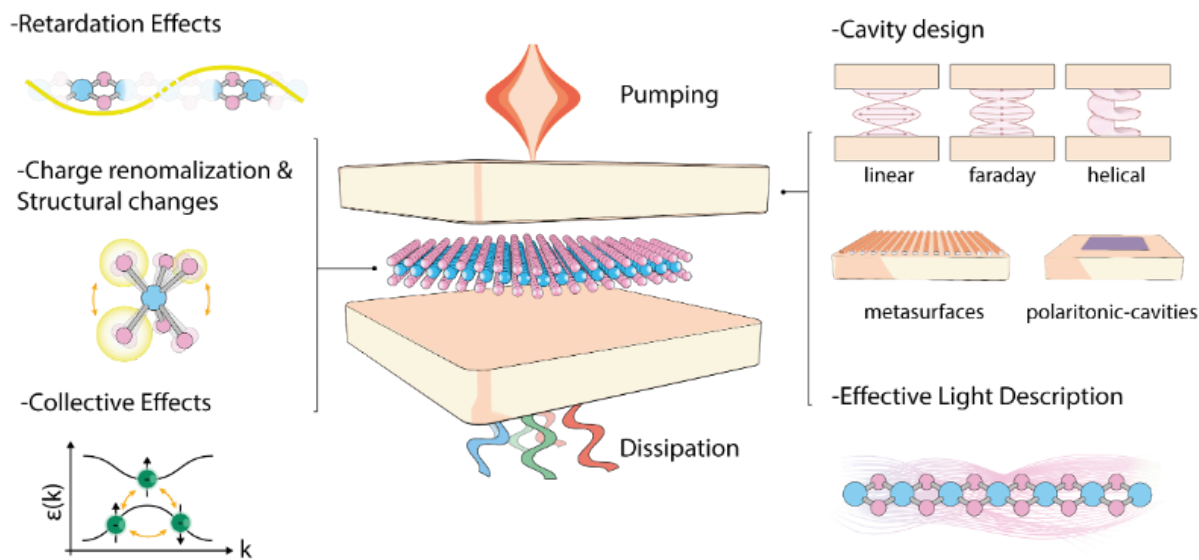

**Figure 1.** Cartoon of a 2D material embedded in a simple Fabry-Pérot optical cavity. Panels to the left showcase the need for an atomistic approach when solving the problem for realistic materials while accounting for retardation, non-perturbative and collective quantum phenomena. Panels to the right depict possible cavity designs, which allow the spatial control of the electromagnetic field beyond the simple Fabry-Pérot cavity. Ultimately, from a practical perspective, we aim at incorporating quantum effects of the background electromagnetic field, including pumping the cavity and inherent dissipation, in an effective way.

**Current and Future Challenges**

Applying QEDFT to cavity materials engineering presents several challenges, both physical and methodological. As illustrated in Figure 1, a successful simulation method should be able to treat complex light-matter phenomena with atomistic accuracy and enable design of realistic cavity environments beyond simple Fabry-Pérot cavities.

From the atomistic point of view, retardation effects, due to the finite speed of light, modify electromagnetic interactions at atomic scales. The multi-way interaction between light and matter in cavities can result in charge renormalization and structural modifications. Additional complexity is added by the collective nature of the light-matter coupling, and hence modelling the connection between microscopic and macroscopic interactions is essential to describe phenomena like polarization-glass transitions [10]. Simulating these effects demands handling many degrees of freedom, highlighting the necessity for efficient theoretical methodologies, such as QEDFT. Quantifying light-matter coupling strengths currently involves fitting experimental Rabi splittings using simplified Hamiltonians that only consider a few electronic states. A predictive model must instead account for the cavity's fabrication and material properties, and the back-reaction of the material within the cavity, i.e., renormalization of the materials screening and cavity mode frequency structure. This feedback influences the coupling and affects both light and matter components.

Beyond the challenges from the constituent particles of the light-matter problem, cavity design itself is crucial in shaping light-matter properties. Most studies focus on simple Fabry-Pérot setups, yet advanced designs such as chiral, helical, and metasurface-based cavities introduce new ways to control materials properties. For instance, chiral cavities may break time-reversal symmetry, enabling equilibrium phase transitions [11], [12]. Moving beyond the long-wavelength approximation (LWA) [13] is another hurdle for modelling advanced cavity designs; recent experiments highlight how spatial gradients in the electromagnetic field impact effects like cavity-modified quantum Hall states [7]. Incorporating these spatial variations is challenging due to different periodicities of photon and matter subsystems.

For QEDFT to become the tool of choice for simulating light-matter systems, it must evolve to serve as a predictive tool. Developing tailored electron-photon exchange-correlation functionals is essential for predicting light-induced phase transitions and related electronic phenomena, and recent velocity-gauge functionals show promise for strongly coupled light-matter systems [14]. Finally, modelling driven systems in pumped cavities, where a few injected photons can alter material phases, requires handling dissipation and complex interactions among electrons, nuclei, and photons.

**Advances in Science and Technology to Meet Challenges**

Paramount for QEDFT is the development of electron-photon exchange-correlation functionals akin to the electron-electron functionals in standard density-functional theory (DFT). Recently, an electron-photon exchange functional suitable for solid-state and extended systems has been developed [14]. Its local-density approximation (LDA) version computes ground states of strongly coupled light-matter systems, and is expected to be further refined through wave-function-based *ab initio* solutions. Similarly to DFT refinements via quantum Monte-Carlo methods, prototypical light-matter systems like the photon-coupled electron gas will help add correlation terms to these functionals [15]. This will allow us to accurately describe how photon vacuum fluctuations affect band structures, electron-phonon coupling, and phase transitions [5].

To address realistic cavity effects, QEDFT can integrate with macroscopic quantum electrodynamics (MQED) approach for lossy photonic environments. A consistent framework combining MQED with (QE)DFT for molecular systems has been developed, though largely within the length gauge [16]. Extending this framework to the velocity gauge would be advantageous for applications in solids. Exploring macroscopic light-matter systems requires incorporating the spatial profile of photon fields and going beyond the LWA. Recent wave-function-based *ab initio* methods, like QED coupled-cluster approaches, show that capturing chiral photon fields requires moving beyond this approximation [17]. In QEDFT, the formulas and approximations beyond the LWA have been theoretically discussed [14]: when the LWA is not applicable, photonic degrees of freedom can no longer be mapped to an electron density but rather the charge-current density [1], [18].

Externally driven cavities introduce complex dynamics among electrons, nuclei, and photons, requiring dynamical treatment of all three subsystems. Coupled QEDFT with Ehrenfest dynamics has been used to simulate finite systems [19], [20] but remains underdeveloped for extended systems. Modelling nuclear dynamics will enable accurate ion relaxation and lattice adjustments, revealing how optical cavities affect crystal structures. Time-dependent modelling capturing dissipation and energy loss into the far field necessitates simulating many photon modes [16] and rewriting it in terms of current-density functional avoids unfavorable computations scalings.

Finally, QEDFT can form the basis of refined *ab initio* methods. Combining QEDFT with Wannier functions can produce precise tight-binding models and improved cavity-dressed Hamiltonians. Extending DFT+U, DFT+DMFT, and related correlated electronic structure methods to QEDFT enables calculating cavity-modified Hubbard U terms for strongly correlated systems with light-matter interactions.

**Concluding Remarks**

The field of cavity materials engineering offers a novel approach to engineer material properties, even at equilibrium, through quantum fluctuations induced by light-matter interactions. QEDFT has emerged as a powerful *ab initio* framework to provide insights into coupled light-matter systems. However, realizing its full potential requires addressing significant challenges. These include moving beyond the LWA to capture spatially varying photon fields necessary to treat realistic cavity, combining QEDFT with MQED to account for lossy electromagnetic environments, and developing charge-current density functionals to accurately describe coupled electron-photon systems in chiral cavity designs. Additionally, extending QEDFT to include time-dependent and driven systems is essential for capturing the complex dynamics of coupled electrons, nuclei, and photons in pumped cavities and in the presence of dissipations. Moreover, experimental validation is crucial to either confirm or challenge theoretical predictions, driving further theoretical and practical advancements in the field. Collaborative efforts between theory and experiment will pave the way for exciting new discoveries in cavity materials engineering.

## Acknowledgements

This work was supported by the Cluster of Excellence 'CUI:Advanced Imaging of Matter' of the Deutsche Forschungsgemeinschaft (DFG) (EXC 2056 and SFB925), the European Union via the Marie Sklodowska-Curie Doctoral Network SPARKLE grant No. 101169225 and the Max Planck-New York City Center for Non-Equilibrium Quantum Phenomena. The Flatiron Institute is a division of the Simons Foundation.

# 8 – Theoretical spectroscopy and topology in the time domain, Floquet states and beyond

**Michael Schüler,** PSI Center for Scientific Computing, Theory and Data, 5232 Villigen PSI, Switzerland and Department of Physics, University of Fribourg, CH-1700 Fribourg, Switzerland.
michael.schueler@psi.ch

**Status**

While angle-resolved photoemission spectroscopy (ARPES) is a standard tool to investigate the band structure of materials [1], progress in generating and controlling ultrafast laser pulses renders time-resolved ARPES (trARPES) one of the most powerful methods to study out-of-equilibrium dynamics of solids [27]. Beyond tracking the relaxation of a system after a moderate-strength pulsed excitation, strong pumps enable the ultrafast manipulation of material properties [2]. One of the most powerful concepts in this field – Floquet engineering – is based on the manipulation of electronic properties by strong periodic driving. In periodic solids, the combined time- and real-space periodicities give rise to Floquet-Bloch states. In analogy to equilibrium band structure, Floquet-Bloch states are characterized by a quasi-energy band structure (see Fig. 1).

The key idea is that the quasi-bands can be different from mere replicas of the equilibrium band structure, featuring gap openings or band crossings of possibly topological origin. Pioneered by Oka and Aoki [3], who showed that the Floquet-Bloch states of graphene driven by circularly polarized light correspond to a Chern insulator, Floquet engineering as a tool to control electronic properties has become a major theme in condensed matter physics, with many theoretical proposals in the literature. A large part of these proposals invoked major simplifications and assumptions, and material-specific predictions that could be observed in experiments were practically non-existing. The situation changed with first-principles methods entering the field. Real-time time-dependent density-function theory (RT-TDDFT) provides a direct way of simulating the time evolution and even simulating the trARPES signal [4]. Another state-of-the-art approach is provided by the Wannier-function method [5]. Besides an efficient interpolation of the Bloch Hamiltonian, the light-matter coupling can be efficiently computed from the Wannier-function method [6]. Hence, the Wannier representation of Floquet- Bloch states can be computed, including the quasi-energy band structure and topology [7].

Despite the many theoretical predictions, experimentally observing Floquet engineering, let alone Floquet-Bloch states, proved to be difficult (see section 6). The picture changed very recently, with a number of successful experiments on a range of material platforms. There is, however, a gap between theoretical predictions and real-world experiments, as including the many effects that become pronounced far from equilibrium – electron-electron and electron-phonon scattering, screening, many-body effects – is challenging. Overcoming these challenges is a crucial cornerstone to understand and utilize light-induced processes in solids, with many potentially game-changing applications in ultrafast electronics and optics.

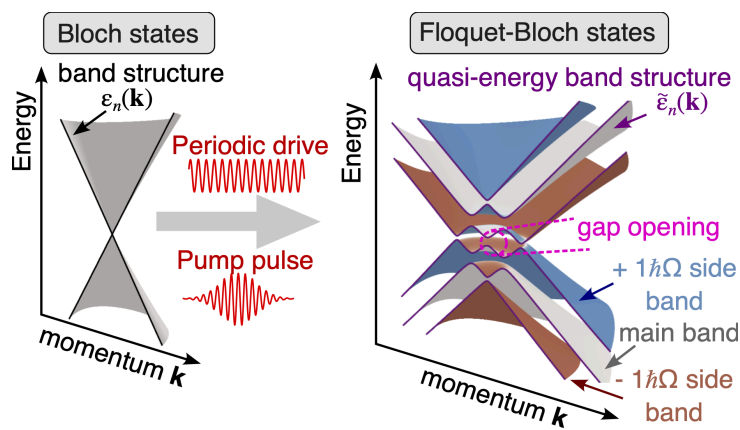

**Figure 1.** From Bloch states to Floquet-Bloch states: applying a time-periodic external drive with frequency Ω to a periodic solid can be treated within Floquet theory in the basis of Bloch basis. The resulting Floquet-Bloch states (here for the example of graphene) are characterized by the quasi-energy band structure. Typical features are the replica bands (here we show the ±1ℏΩ replica) as well as a renormalization of the bands and gap openings. In practice, the signatures of Floquet-Bloch states can also be observed for pulses comprising sufficiently many optical cycles.

**Current and future challenges**

RT-TDDFT provides a parameter-free prediction of electron dynamics in systems containing hundreds of atoms. Even the simulation of trARPES and the signatures of Floquet-Bloch states in two-dimensional (2D) materials is possible [4]. The major drawback of the RT-TDDFT approach to trARPES is the enormous computational demand, due to the necessity to simulate the system along with a large portion of vacuum. The direct simulation of surfaces of bulk materials or 2D systems on bulk substrates is currently out of reach. Since ARPES is a particularly surface-sensitive technique, the description of a solid including its surface and the surface dielectric response is paramount for understanding the manifestation of Floquet-Bloch states in experiments. The time-dependent nonequilibrium Green's function (td-NEGF) approach provides an alternative way of describing excitations in pump-driven solids. The central quantity, the one-body Green's function $G^<(k, t, t')$, not only delivers the density matrix, but also the gauge-invariant trARPES spectrum [9]. The key difficulty lies in computing the two-time dependence of $G^<(k, t, t')$, due to the enormous numerical effort of solving the underlying Kadanoff- Baym equations [10]. Simplifications are inevitable to make progress; the generalized Kadanoff-Baym ansatz (GKBA) [11] has proven particularly useful in this context [12]. The still considerable computational effort renders full-fledged first-principles simulations of ultrafast dynamics including electron-electron scattering impractical at present.

The second major ingredient to simulating trARPES are the photoemission matrix elements. Not only are matrix elements important for interpreting ARPES experiments, but information about the quantum-geometrical properties, such as orbital angular momentum (OAM) and the Berry curvature, is contained as well [13]. In equilibrium ARPES, extracting information on the OAM of Bloch electrons by using circularly polarized probe pulses has become an active field [14,15,16]. Extending the concept of measuring the OAM out of equilibrium would provide access to light-induced topological states and other nonequilibrium protocols [17,18].
Overcoming the methodological and computational challenges explained above will greatly facilitate theoretical photoemission spectroscopy in and out of equilibrium. Predictive power for simulating the trARPES signature of (topological) Floquet-Bloch states is a key step to guiding the at present very challenging pump-probe experiments. Because of the experimental challenges, successful Floquet engineering has only been realized for a select few materials. Drawing the analogy with the field of materials design, computational pump-probe spectroscopy would open the door to nonequilibrium materials design. Besides finding optimal pump parameters, the specific material properties that enable creating and controlling Floquet-Bloch states are unclear. For instance, it was believed that Floquet engineering is not possible in metallic systems due to the fast decoherence [8]. In recent experiments [19,20] Floquet-Bloch states in semimetallic graphene were clearly observed, indicating that the Floquet physics is robust. The role of scattering mechanisms is, however, far from understood. Another key ingredient to control the electronic structure by laser driving is the light-matter coupling itself. Here the challenge is of conceptual rather than computational nature: under which conditions are strong inter-band hybridizations possible? Questions of the specific nature of light-matter coupling in solids were discussed more recently in the context of the quantum geometry of Bloch electrons [21]. The quantum geometric perspective establishes a link between band topology and light-matter interaction in solids, and also provides new insights into nonlinear spectroscopy [22,24,25].

**Advances in science and technology to meet the challenges**

There are promising developments in simulating ultrafast phenomena in solids that will likely provide a new level of predictive power for Floquet engineering and strong-field phenomena in general. RT-TDDFT is among the most powerful *ab initio* methods (some recent advances are covered by other contributions to this roadmap). Connecting to Floquet spectra is, in principle, straightforward by analyzing the time-dependent Kohn-Sham Hamiltonian in the Floquet language (see section 6). Connecting to trARPES is difficult, since the Green's function needs to be computed. The projection of the time-dependent Kohn-Sham states onto the equilibrium bands or a Wannier basis is a promising approach that would provide the Green's function and thus enable trARPES simulations.

If many-body effects play an important role, methods beyond (standard) RT-TDDFT need to be employed. The td-NEGF machinery established itself as one of the most promising paths forward. For instance, the topological Floquet-Bloch states in graphene in the presence of electron-electron and electron-phonon scattering have been studied in ref. [17], combining a tight-binding model, real-time dynamics on GKBA level, and realistic photoemission matrix elements. As a key finding, the circular dichroism in trARPES proved to be a robust fingerprint of the light-induced topology. The first steps to extending this nonequilibrium machinery to first-principles level have already been made. For instance, the signature of light-induced band hybridization in bulk black phosphorus in trARPES were simulated in ref. [23] based on the *ab initio* Wannier function approach (albeit without scattering effects), and excellent agreement with experiments was obtained (see Fig. 2).

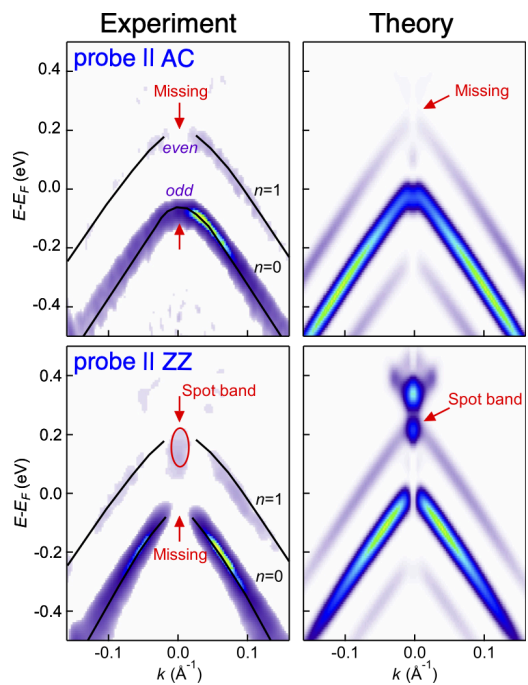

*Figure 2.* trARPES signal of the Floquet state in bulk black phosphorus from experiments and theory. The pump pulse is polarized along the armchair (AC) direction, while the probe light is oriented either along AC or zigzag (ZZ) direction. While for both cases the $n = 1$ side band of the top valence can be observed, a "spot band" feature appears for ZZ- polarized probe. The spot band is a manifestation of light-induced hybridization of valence and conduction band, which is confirmed by theory and trARPES simulations. Adapted from ref. [23].

With accurate modeling of the light-matter coupling available, the remaining challenge is to advance the simulation of ultrafast many-body dynamics itself. Recent developments in the field are indeed promising. For instance, the td-NEGF machinery has been pushed to a new level, including linear-scaling GKBA (see section 3). From the first-principles perspective, the time-dependent adiabatic $GW$ [26] has been demonstrated to capture key many-body effects such as excitons (see section 2).

**Concluding remarks**

The community is in an excellent position for taking the next step towards predictive first-principles methods that would expand computational materials design to the ultrafast domain. Not only would efficient and robust on-demand ultrafast band engineering become a reality, but theoretical progress in ultrafast dynamics and spectroscopy would also greatly benefit ultrafast materials science in general, where potential break-through applications in ultrafast optoelectronics or information processing are likely to be found.

**Acknowledgements**

M.S. acknowledges support from SNSF Ambizione Grant No. PZ00P2-193527. This research was supported by the NCCR MARVEL, a National Centre of Competence in Research, funded by the Swiss National Science Foundation (grant number 205602).

## 9 – Ab initio time-resolved spectroscopy

**Davide Sangalli,** Istituto di Struttura della Materia (ISM - CNR), Area della Ricerca di Roma 1, Monterotondo Scalo, Italy.
davide.sangalli@ism.cnr.it

**Claudio Attaccalite,** CNRS/Aix-Marseille Université, Centre Interdisciplinaire de Nanoscience de Marseille UMR 7325 Campus de Luminy, 13288 Marseille cedex 9, France.
claudio.attaccalite@univ-amu.fr

**Myrta Grüning,** School of Mathematics and Physics, Queen's University Belfast (BT7 1NN University Road, Belfast Northern Ireland).
m.gruening@qub.ac.uk

**Status**

In time-resolved spectroscopy (TR-S), a system is first excited by a pump laser, and then probed at a given delay time with different techniques: TR-absorption (abs) [1, 2, 3], TR-Second Harmonic and High Harmonic Generation (SHG/HHG)[4], TR-Angular Resolved Photo-Emission Spectroscopy (ARPES) [5], electron or X-ray TR-diffraction [6]. With the addition of more laser pulses, techniques such as Four-Wave Mixing, 2D spectroscopy, or Pump-Push&Probe experiments can be performed. TR-S allows the study of correlated electron-photon-ion dynamics in advanced materials on extremely short time-scales. However, the interpretation of these experiments is very challenging due to the variety of physical processes involved.

ab initio approaches, which are routinely used to predict structural and electronic properties of advanced materials, are emerging as a valuable tool to interpret and guide these experiments. Time-dependent density-functional theory (TDDFT) was one of the first tools used by the condensed-matter community. The real-time (RT) implementation is naturally suited to describe nonlinear couplings between electrons and laser pulses and has been employed for 20 years to model TR-S experiments [7]. RT-TDDFT for extended systems is nowadays available in many standard ab initio codes such as Octopus [8], Salmon [9], Yambo [10], Exciting [11], and Elk [12], with more codes under development.

RT-TDDFT's key advantage is its simplicity (within standard exchange-correlation approximations) but this also limits its application as it misses excitonic, relaxation, and dissipation effects. To properly account for excitonic effects, ab initio approaches based on nonequilibrium Green function (NEGF) have been developed [13], within the static Hartree + Screened Exchange (HSEX) approximation [10].

The coupling between atomic and electronic dynamics can be treated at different levels of approximation. For example, Ehrenfest dynamics captures the coupling with classical atomic motion [14], while nonadiabatic effects, such as relaxation and dissipation processes, require dynamical approximations within NEGF. Early attempts to make ab initio NEGF simulations with dynamical approximations feasible required extensive simplifications [15], leading to TD-HSEX equations coupled with rate equations for electronic occupations [16]. More advanced approaches have so far relied on simplified, not fully ab initio methods [17].

A further challenge is the modeling of the detection process. This has either been described directly, by including the delayed probe signal in RT simulations, or by reconstructing the measured signal from nonequilibrium quantities. [18] For TR-abs one can input nonequilibrium populations in the excitonic Hamiltonian [19] while the nonequilibrium density-matrix can be used to reconstruct the TR-ARPES signal [20]

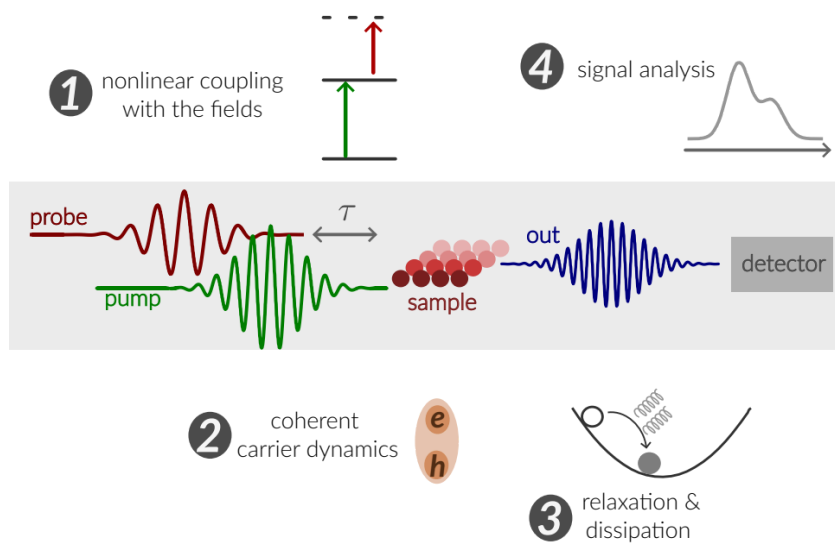

.

**Figure 1.** Scheme of a time-resolved spectroscopy experiment: the pump excites carriers in the system. a second laser beam at some time-delay τ probes the carriers' dynamics. Simulations of time-resolved spectroscopy require to consider (1) the nonlinear coupling with the field and to capture both (2) the coherent carrier dynamics—including relevant many-body effects—and (3) the dephasing relaxation and dissipation phenomena. Finally, (4) the transient signal must be extracted from the simulation output and analysed.

**Current and Future Challenges**

Bringing ab initio approaches in the nonequilibrium regime calls for accurately describing the nonlinear electron-laser couplings, correlated coherent dynamics, dephasing relaxation and dissipation, and the detection process (see Fig. 1).

Nonlinear couplings: (intense) laser pulse can be modeled either via the minimal coupling approach [21], using a Berry phase formulation [22], or computing dipoles with Wannier functions [23]. Each of these techniques have advantages and limitations when combined with TDDFT or NEGF simulations.

Correlated coherent dynamics results from electron-electron (*ee*) and electron-nuclei (*en*) interactions. Within RT-TDDFT, adiabatic *ee* interactions are captured within local approximations, while including on-site correlation [24] or long-range exchange [25] is challenging. TD-HSEX includes long-range correlations but at a greater numerical cost. Classical *en* coupling, within Ehrenfest dynamics, is available in RT-TDDFT, while approaches to combine it with TD-HSEX are still under development [26].

Modeling dephasing relaxation and dissipation, remains the greatest challenge. This requires accounting for nonadiabatic interactions and dynamical many-body self- energies. Dynamical self-energies are also needed to account for nonequilibrium screening. The main difficulty is the computational cost, which grows rapidly with the system size. To date, the ab initio community has focused on accurately calculating phonon- and photon-induced scattering, as well as electron and exciton lifetimes and electron dephasing times from these equations [27, 28, 29]. This area remains an active research field with ongoing numerical and computational challenges [30].

Accurately modeling the detection process presents several ongoing challenges. It requires incorporating the correct physics to capture energy shifts [19, 1] and changes in electronic [3, 5] and structural [31, 2] properties commonly measured in TR-S experiments. Additionally, addressing the probe laser pulse's effect is essential for accurately interpreting coherent signals in TR-S. A recent and exciting challenge involves modeling TR-SHG/HHG signals [4], where coherence between the pump and probe must be precisely accounted for.

**Advances in Science and Technology to Meet Challenges**

Although the above challenges are common to all models of TR-S, the ab initio framework poses a further challenge due to the high, sometimes prohibitive, computational cost that comes from the parameter-free, material-specific treatment of electron systems. In recent years, significant progress in science and technology has been directed towards addressing this challenge. The increase in computing power with newly available High Performance Computing (HPC) centres, and the development of ab initio codes for exascale computing are opening the way to simulations which were not feasible until a few years ago. In particular, extensive work on parallelization (MPI and OpenMP) and porting some of the codes to GPUs has allowed more complex materials to be simulated with better approximations. This progress has been possible thanks to the support of large projects such as the European Centers of Excellence MaX [32] and NOMAD [33], and similar initiatives in the United States.

However the tuning of ab initio codes has, thus far, focused on more established simulations, e.g. on DFT ground state methods or equilibrium spectroscopy approaches. The optimization of codes for RT simulations is still in its infancy, and there is plenty of room for technical improvements. Moreover, in two cases, realistic modeling of TR-S often exceeds the capabilities of ab initio simulations, even on HPC machines: (I) when accounting for correlation and dissipation, especially in the coherent regime, and (ii) when incorporating environmental effects alongside complex materials and structures. Significant progress has been made in tackling the latter challenge both in the ground state and for response functions [34, 35]. However, no approach is yet available for including environmental effects into real-time electron dynamics within the Green function formalism. Possible solutions for tackling the former challenge include synergies with simplified models. Software optimization, e.g. MPI/OpenMP/GPU porting, that focused so far on ab initio codes, must be extended to hybrid ab initio-modellistic codes, so to address either the transition from the coherent to the incoherent regime and dynamical screening effects.

**Concluding Remarks**

We briefly discussed how developments of TDDFT and NEGF within the ab initio domain, together with the advancement in HPC, are opening the way to model nonequilibrium dynamics and TR-S. The field is still relatively new. On one hand, approaches to model coherent electron dynamics are available or under development, but the codes still need to be optimized to take advantage of HPC to model advanced materials. On the other hand, the main limitation remains the description of the dynamical correlations responsible for the transition from the coherent to the non-coherent regime. This description involves the coupling of the electron dynamics with other degrees of freedom, such as the atomic motion or the environment. Finally, there is the need to introduce the coupling with the environment in ab initio schemes in an effective way, or, more generally, by looking for synergies with approaches based on models, in such a way as to threaten more realistic systems. There is room for improvement, and a close connection with the experimental community will be critical for properly addressing the theoretical and computational developments.

**Acknowledgements**

We acknowledge support from the Doctoral Network TIMES, funded under the call HORIZON-MSCA-2022-DN-01 (Grant Agreement No 101118915). D. S. acknowledges MaX "MAterials design at the eXascale", co-funded by the European High Performance Computing joint Undertaking (JU) and participating countries (Grant Agreement No. 101093374), and the Innovation Study Isolv-BSE, which has received funding through the Inno4scale project, funded by the European High-Performance Computing Joint Undertaking (JU) (Grant Agreement No 101118139). C. A. acknowledges the ANR project COLIBRI No. ANR-22-CE30-0027, and support from the Excellence Initiative of Aix-Marseille Université - A*Midex, thought the INDIGENA project.

# 10 – Ab initio theoretical description of ultrafast magnetization dynamics in quantum materials

**S. Sharma,** Max-Born-Institut für Nichtlineare Optik und Kurzzeitspektroskopie, 12489 Berlin, Germany. Institute for theoretical solid-state physics, Freie Universität Berlin, Arnimallee 14, 14195 Berlin, Germany
Sangeeta.Sharma@mbi-berlin.de

**S. Shallcross** Max-Born-Institut für Nichtlineare Optik und Kurzzeitspektroskopie, 12489 Berlin, Germany
Samuel.Shallcross@mbi-berlin.de

*Keywords: ab initio simulations, ultrafast magnetization dynamics, quantum materials*

**Status**

Ultrafast laser pulses induce a remarkably rich cascade of processes in condensed matter, invoking, as time evolves from the initial excitation, successively more degrees of freedom on increasing spatial length scales [1-3]. The present day state-of-the-art theory is well established only at the ultimate spatial and temporal limits of this excitation process; at femtosecond time scales the dynamics of free carriers can be treated with quantitative accuracy by *ab initio* time dependent density functional theory (TD-DFT) [1,3,6,8,13] while, at the other limit of long time scale, models and quasi-classical theories address the decoherent diffusive regime of light induced currents and long-time evolution of complex magnetic textures and structures.

Light induced changes in magnetic order present a rich early time tapestry of nonequilibrium physics, in which *ab initio* theory has not only aided the interpretation of experiment [4] but predicted new time-dependent phenomena subsequently confirmed by experiments. One notable example is optical intersite spin transfer [2], in which magnetic matter undergoes ultrafast demagnetization or rearrangement of local structure (for example from anti-ferromagnetic to ferromagnetic). Beyond the dynamics of free carriers, the inclusion of excitonic physics [10] and lattice dynamics [7,14,18] on an equal footing is presently under development (see sections 3 and 4), ultimately promising a theoretical scheme in which the light-induced dynamics of phonons, excitons, magnons, and free carriers can be predicted and thus controlled by designed light pulses. Laser pulses further allow control over crystal momentum in solids, an entirely non-perturbative effect of ultrafast light-matter interaction, and a feature that can be exploited for remarkable control over spin- and valley-tronics [16].

*Ab initio* simulations of key experimental techniques in transient spectroscopy, e.g., magneto-optical Kerr effect (MOKE), x-ray magnetic circular dichroism (XMCD), or angle- resolved photoemission spectroscopy (ARPES), have led to remarkable agreement with experiment for the temporal evolution of local moments in complex alloys [5] and light-induced changes in band structures [13] (see sections 9 and 10). The ultrafast regime of light-matter interaction in condensed matter thus both contains a wealth of unexpected phenomena as well as established tools that permit the testing of predictions in experiment.

**Current and Future Challenges**

It is in the borderlands between the two temporal extremes of ultrafast and quasi-classical that key theoretical challenges remain. At the simplest level this is the "joining up" of *ab initio* and quasi-classical domains by, for example, calculating the parameters of an effective Heisenberg model from an underlying nonequilibrium electronic state, in this way connecting not only distinct temporal domains but also spatial domains: the unit cell that describes the nonequilibrium electronic structure with the large length scale physics that can be addressed by the Heisenberg model. Similar approaches can be envisaged in other contexts: (i) *ab initio* forces calculated post-pulse, but still at femtosecond times, providing the initial conditions for molecular dynamics simulation to examine angular momentum transfer and light-triggered structural changes; (ii) *ab initio* occupations and currents as the input for quasi-classical treatments such as those based on Boltzmann transport theory.

However, the existence of such simple "horizons" at which a fully quantum theory can be "joined up" to emergent model variables is ultimately doubtful. An alternative technique is to either selectively enhance or remove degrees of freedom from the *ab initio* approach. The former, the route of the "long-range ansatz" established for the ground state [10], allows both the unit cell scale as well as a second distinct large length scale, e.g., that of domain walls or skyrmions, to be included in the simulation. The second approach, removing degrees of freedom, consists in replacing elements of a nanostructure that, e.g., serve to generate a current by appropriate boundary conditions on the "active" part of the device [16].

While the material-dependent pathways in which effective classical or quasi-classical variables emerge as stable structures from the "big bang" of ultra-fast excitation remain unclear, a fundamental issue that is only beginning to be addressed is the role of quantum decoherence in this transition. This represents a situation of a different character: decoherence times fall in the range of 5-100 fs and so well within the domain of *ab initio* physics, yet all *ab initio* theories are solutions of the Schrödinger or Dirac equations and so are fully coherent. While attempts to introduce a version of decoherence into *ab initio* theory have been made via density matrices [9], it remains the case that the division, central to the treatment of decoherence, of the physics into "bath" and "system" degrees of freedom is ultimately antithetical to the *ab initio* philosophy.

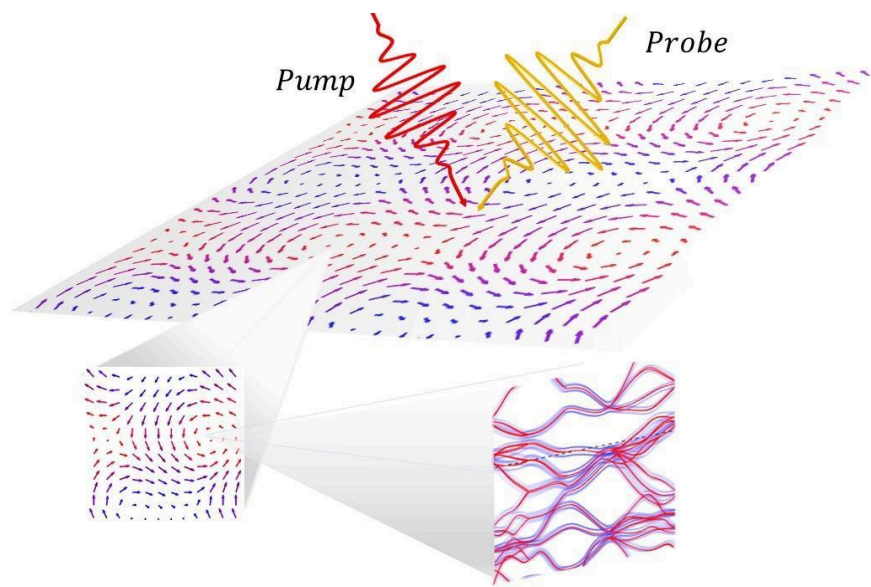

**Figure 1** *Schematic showing multiscale approach to study pumped long length scale spin textures; unit cell length scales of few Angstrom to ultra long length scale of the texture of a few nano-meters.*

**Advances in Science and Technology to Meet Challenges**

The challenges implied by the extension of theoretical methodologies to capture the processes of ultrafast excitation in condensed matter can broadly be divided into two schemes: the "meshing together" of distinct techniques applicable to different temporal and spatial domains, and the development of *ab initio* theory itself to capture physics beyond the unit cell and femtosecond scales it is now well established for. To accomplish this will involve both the creation of new theoretical tools, but also the answering of deep fundamental questions, e.g., concerning the nature of decoherence in the solid state.

(1) *Extension of the long-range ansatz to the time domain* would allow the *ab initio* treatment of the dynamics of spin and charge densities on large length scales; a scheme of this multi-scale approach is shown in Fig. 1. This would permit investigation of light control of, e.g., skyrmion textures, spin domain walls, and spin shock waves, but not large scale structures such as heterostructure twist geometries.

(2) *ab initio theory as a basis for initial conditions in molecular dynamics*. Early-time dynamics of electrons and the underlying lattice structure, treated via the Ehrenfest approach [15], providing the initial conditions for molecular dynamics with the boundary time fixed by the ceasing of the dynamics of excited state occupations. Such an approach would allow investigation of light-triggered delamination of 2d materials.

(3) *TD-DFT and quasi-classical transport*. *ab initio* calculation of ultrafast light-induced charge and spin currents in an open system scheme [9] as an initial condition for longer time quasi-classical calculation of current via a Boltzmann transport approach.

(4) *Inclusion of decoherence*. A fundamental question is the role of disorder self-averaging leading to loss of phase structure within a large unit cell – physics that can in principle be described in a coherent calculation – versus quantum decoherence, i.e., entanglement with bath freedoms, for which a density matrix is essential.

(5) *Extension to long times* can proceed either via (i) *ab initio* dynamics providing initial conditions for effective models such as tight-binding or Heisenberg, but also (ii) by providing a "non-frozen" Wannier scheme in which the tight-binding parameters are updated to incorporate dynamical effects of the excited state electronic structure.

**Concluding Remarks**

The present day state of the art allows theory to predict and partner experiments in a comparatively circumscribed set of domains: *ab initio* theory is restricted to times on the order of hundreds of femtoseconds and scales on the order of the unit cell; effective-variable models, while able to access spatial and temporal domains beyond those of *ab initio* theory, are underpinned by a choice of model assumptions and so intrinsically lack the predictive power of *ab initio* theory. Key future challenges are therefore (i) to extend the domain of *ab initio* theory via selection of the key variables of the dynamical processes induced by light, and (ii) to employ *ab initio* theory as a route towards providing the initial conditions and parameters of effective-variable models.

**Acknowledgements**

*Sharma would like to thank DFG for funding through TRR227 (project A04) and Leibniz Professorin Program (SAW P118/2021). Shallcross would like to thank DFG for funding through SH 498-7/1.*

## 11 – Real-time simulations of nonequilibrium phonon dynamics

**Fabio Caruso,** Institute of Theoretical Physics and Astrophysics, University of Kiel, Germany
caruso@physik.uni-kiel.de

*Keywords: Electron-phonon interactions, ultrafast dynamics, phonon dynamics*

**Status**

Following electronic excitation with ultrafast pulses, the nonequilibrium dynamics of the crystalline lattice is governed by the energy flow between electronic and vibrational degrees of freedom. These interactions give rise to a variety of emerging phenomena, including the formation of nonequilibrium phonon populations [Sei21], coherent phonons [Hei20], phonon damping and softening, band-structure renormalization [Hei20], structural and topological phase transitions [Sie19], and polaron formation. These processes are of fundamental scientific interest, but also offer new opportunities for transient control of materials properties on ultrafast timescales [Man17]. The ability to directly probe the lattice dynamics and electron-phonon interactions in real time has been boosted by recent advances in experimental ultrafast science [Ger17], as exemplified by the developments of ultrafast electron diffuse scattering [Fil22] and frequency-domain angle-resolved photoemission [Hei20]. These techniques have enabled the exploration of novel transient phenomena resulting from the lattice dynamics, enhancing the urgency of developing transferable theoretical models for phonons out of equilibrium.

From an ab initio materials modelling perspective, describing the lattice dynamics requires to accurately capture the interplay of electron and lattice motion in presence of electron-phonon interactions in driven solids. The time-dependent Boltzmann equation (TDBE) is a powerful theoretical framework to address this challenge [Car21]. It models the evolution of electron and phonon distribution functions ($f_{nk}$ and $n_{q\nu}$, respectively) through a set of coupled integro-differential equations – in the form $\partial_t f_{nk} = \Gamma_{nk}[f_{nk'}, n_{q\nu}]$ and $\partial_t n_{q\nu} = \Gamma_{nk}[f_{nk'}, n_{q\nu}]$ – where electron-phonon scattering and other interactions can be incorporated via the collision integrals $\Gamma_{nk}$ [Car21,Ton21].

Early applications of the TDBE emerged in the early years of ultrafast science, and the advancement of computational materials modelling have recently enabled fully ab initio simulations based on the TDBE [Jha17]. Recent applications include high-field transport [Mal21], valley-polarization dynamics in two-dimensional semiconductors [Xu21, Pan24], modelling of ultrafast diffuse scattering [Pan23], and coherent-phonons in pump-probe photoemission [Eme24]. These studies have been enabled by the development of computational workflows (Fig. 1) which integrated the TDBE formalism into highly-parallel codes for ab initio calculations and electron-phonon coupling [Giu17]. Most importantly, TDBE-based simulations are emerging as a powerful tool for theoretical spectroscopy in real time, enabling the accurate description of nonequilibrium phenomena resulting from the emergence of nonequilibrium phonon populations and coherent phonons, thereby providing a valuable tool for the interpretation of pump-probe experiments (Fig. 2).

As the field progresses, concerted efforts are essential to push the boundaries of both computational and theoretical methods. This contribution highlights key challenges and emerging directions that must be addressed to further enhance the modeling capabilities of ab initio methods based on the TDBE approach.

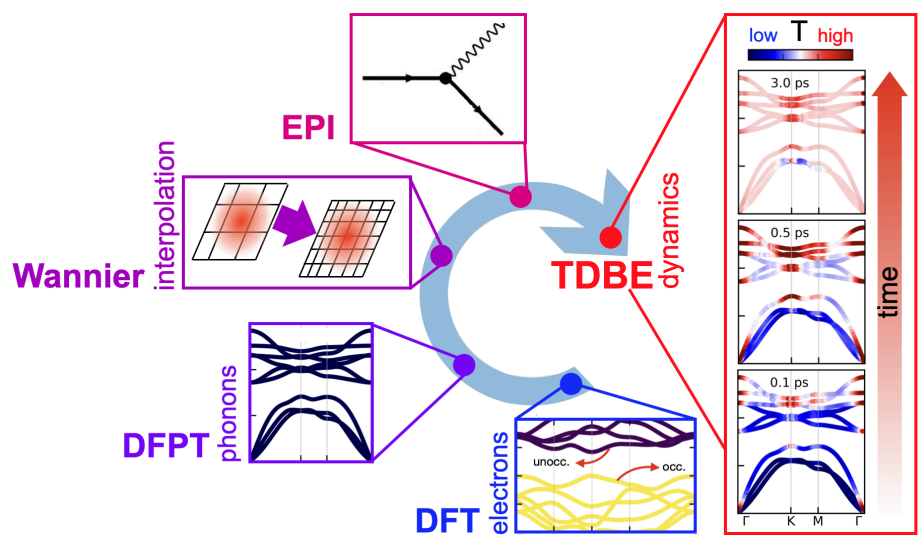

**Figure 1.** Computational workflow for real-time simulations of the non-equilibrium phonon dynamics based on the TDBE formalism. Adapted from [Car21b].

**Current and Future Challenges**

*Computation challenges.* Ab initio simulations based on the TDBE formalism have thus far focused on a narrow selection of 2D materials and simple bulk solids. The key factor hindering wider applications is computational efficiency: convergence requires ultra-dense k-point meshes (typically exceeding $10^5$-$10^6$ points); additionally, time propagation for $10^4$-$10^5$ time steps is required to approach the timescales relevant for electron-phonon and phonon-phonon scattering (5-10 ps); computation and storage of coupling matrix elements on dense grids further represent major bottlenecks. Devising new strategies to tackle this complexity is a prerequisite to extend the applicability of TDBE simulations. Importantly, these problems are common to high-level approaches (e.g., density matrix and nonequilibrium Green's functions). Thus, finding a solution to these issues marks a pivotal challenge for advancing the field of ab initio ultrafast science as a whole.

*Theory limitations.* Describing electron and nuclear dynamics through distribution functions entails several limitations on the ability of the TDBE to address complex dynamical phenomena. Specifically, it fails to capture electronic coherence arising throughout photoexcitation and effects arising from a correlated electronic ground state. Additionally, a description of anharmonic effects beyond third-order phonon-phonon scattering still remains unexplored, and light-induced structural changes and band-structure renormalization are entirely absent, as they cannot be captured in a perturbative framework. This highlights that the strengths of the TDBE are limited to capturing population dynamics and thermalization within a perturbative Markovian approximation, and it cannot be straightforwardly applied in cases of structural instabilities or to the description of other non-perturbative phenomena. Yet, the development of ad-hoc extensions of the formalism or its integration with higher-level theories could substantially help expand the TDBE's modelling capabilities.

*Need for Validation.* Establishing protocols for benchmarking and validating ultrafast dynamics simulations is becoming increasingly critical. The availability of robust benchmarks and data sets, consisting of high-level theoretical and experimental data is essential for ensuring the accuracy, reliability, and predictive power of new methods and models focusing on complex ultrafast phenomena (see sections 13 and 14).

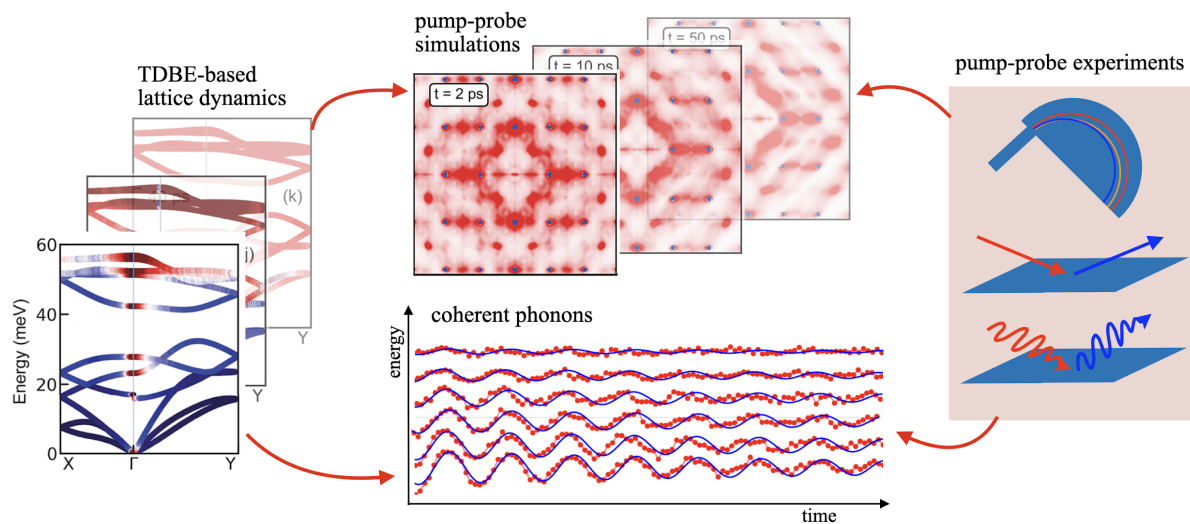

**Figure 2.** Theoretical spectroscopy of non-equilibrium phenomena based on the TDBE formalism. Lattice dynamics simulations within the TDBE formalism are emerging as powerful tools to probe and understand the influence of coherent phonons and non-equilibrium phonon populations in pump-probe experiments. Adapted from [Sei21].

### Advances.

Advancing numerical techniques for TDBE simulations is a high-priority requirement for ab initio studies of ultrafast processes at picosecond timescales. Recent efforts in this direction have focused on forecasting the nonequilibrium behavior of electron and phonon distribution functions via dynamic mode decomposition [Mal24], and on the acceleration of computation of electron-phonon matrix elements. Future developments could further largely benefit from the integration of novel computing paradigms (such as GPU acceleration, machine learning, tensor decomposition) into existing implementations. Gaussian processes, tetrahedron integration for the momentum integrals [Kaw14] exemplify unexplored opportunities with a strong potential of accelerating the computation of collision integrals and time propagation.

Recent developments have extended the TDBE framework to enable the description of coherent phenomena with minor computational overhead. The semiconductor electron-phonon equations [Ste24], for example, generalize the TDBE to account for electron and vibrational coherence (see also section 12). An additional promising route to capture electron coherence consists in interfacing the TDBE with a density-matrix or nonequilibrium Green's function formalism for time scales shorter than 100 fs (sections 2 and 3) – based e.g., on time-dependent adiabatic GW [Cha23]. To address the description of coherent phonons and the ensuing electronic-structure renormalization, time-propagation of the TDBE can be combined with the lattice equation of motion and with a perturbative treatment of electronic coupling to coherent phonons [Eme24]. To further enable a realistic description of light-induced phase transitions and metastable structural changes, future developments should focus on extending the treatment of lattice motion (coherent phonons) and electron-phonon interactions in presence of lattice anharmonicities beyond third order. By leveraging these concepts, the modeling capabilities of the TDBE can be significantly extended without drastically increasing numerical costs.

### Concluding Remarks

The TDBE and its extensions are emerging as essential theoretical tools for modeling ultrafast phenomena, and particularly for capturing nonequilibrium phonon dynamics across timescales ranging from the onset of electron-phonon scattering (100 fs) to full thermalization (50 ps). Present and future advancements in this field hold the promise of expanding its numerical applications to a broader range of materials and more complex dynamical scenarios that are currently beyond reach. Achieving these goals will be an important step to advance computational modelling of ultrafast phenomena as a whole, establishing new tools for theoretical spectroscopy out of equilibrium.

## Acknowledgements

*F.C. acknowledges funding by the Deutsche Forschungsgemeinschaft (DFG), projects number 443988403 and 499426961. Discussions with Yiming Pan are gratefully acknowledged.*

## 12 – Theory of Nonlinear and Chiral Phononics

**Dominik M. Juraschek,** Department of Applied Physics and Science Education, Eindhoven University of Technology, Eindhoven, Netherlands
d.m.juraschek@tue.nl

**Michael Fechner,** Max Planck Institute for the Structure and Dynamics of Matter, Hamburg, Germany
michael.fechner@mpsd.mpg.de

**Status**

Optical phonons provide a unique route to dynamically controlling the properties of materials, because their electric dipole moments directly couple to the electric field of light. The advent of high-field terahertz and mid-infrared pulses in the past decade has enabled resonant driving of optical phonons with large amplitudes – so large that the harmonic approximation breaks down and nonlinear couplings to other phonons, collective excitations, and degrees of freedom become important. A new field of "nonlinear phononics" was born, in analogy to nonlinear optics, where phonons replace photons in the scattering processes [1,2]. Early work focused primarily on enhancing and inducing superconductivity, whereas today, control of various electronic phases has been demonstrated, including magnetism and ferroelectricity [3]. Further, the generation of high-field circularly polarized pulses has enabled exciting circularly polarized, or chiral, phonons that carry angular momentum, leading to the field of "chiral phononics". The rotational motion of the ions around their equilibrium positions produces effective magnetic fields, similar to an electromagnetic coil, that couples to and controls magnetic order [4].

While nonlinear and chiral phononics have fundamentally been enabled by developments in ultrafast radiation sources in the frequency range of optical phonons, typically between 1-40 THz, initial theoretical predictions preceded them by several decades. From a theoretical perspective, the most prominent ones can be grouped into three categories, illustrated in Fig. 1: (a) Phononic rectification, in which driven phonons create a unidirectional force that rectifies the vibrational motion of the atoms along the eigenvectors of another coupled phonon [2]. The so-induced structural distortion changes the geometry and distances between the atoms in a material and hence the electronic interactions responsible for its properties. (b) Frequency conversion, in which energy is coherently transferred between the nonlinearly driven phonons and other phonons and collective excitations. These mechanisms include ionic Raman scattering [2,5], sum-frequency excitations [6], and parametric excitations [7]. (c) Chiral phonomagnetic fields, in which circular motions of the atoms in a material produce real and effective magnetic fields through charge currents and coupling to electronic angular momentum [8], [9]. These processes are typically investigated through time-dependent mode-coupling models or quasi-steady-state (Floquet) methods [10]. However, the description of nonlinear and chiral phononic processes in materials faces some challenges, which we will outline in the following.

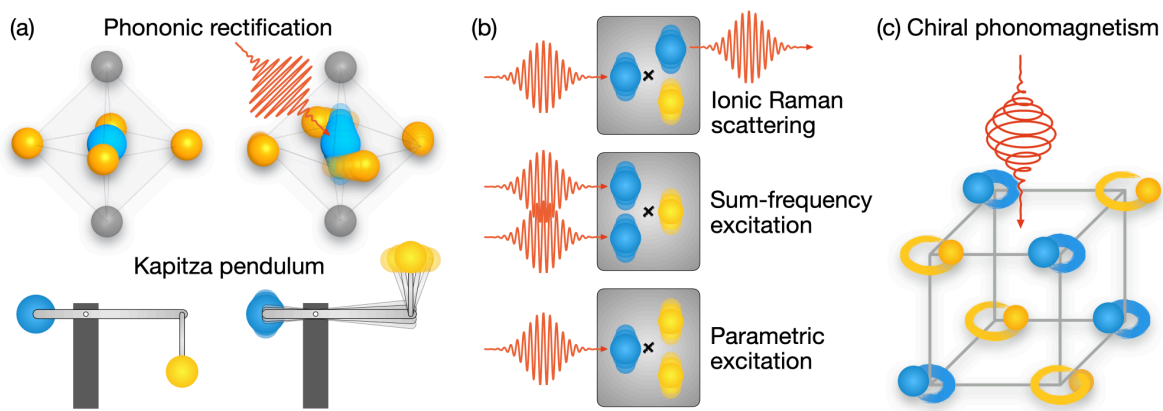

**Figure 1.** *Nonlinear and chiral phononic mechanisms*. (a) Phononic rectification, in which the atoms are displaced along the eigenvectors of a phonon mode (yellow motion) nonlinearly coupled to a driven phonon mode (blue motion). The Kapitza pendulum (bottom panel) is a classical analog, in which the equilibrium position of a pendulum is inverted by a nonlinear driving force. (b) Frequency conversion mechanisms: Ionic Raman scattering, in which a driven phonon mode (blue) scatters by and creates a nonlinearly coupled mode (yellow). Sum-frequency excitation, in which two driven phonon modes are converted to a third, nonlinearly coupled one. Parametric excitation, in which a driven phonon mode downconverts to two nonlinearly coupled ones. (c) Chiral phonomagnetism, in which a circularly polarized pulse drives circular motions of the atoms, producing effective magnetic fields reminiscent of atomistic electromagnetic coils.

**Current and Future Challenges**

*Superconductivity*: Being one of the first electronic orders to be studied with nonlinear phononics, various theories have tried to explain the enhancement of the superconducting temperature in various high-Tc materials, such as YBCO and $K_3C_{60}$. Original explanations involved structural displacements induced by phononic rectification that favor coherent transport. More recent studies considered parametric amplifications of electron-phonon coupling and the associated pairing mechanisms [11,12]. While ongoing experiments are making significant strides toward a definitive evidence of light-induced superconductivity—such as detecting the ultrafast Meissner effect [13]—a comprehensive theoretical framework to satisfactorily explain these phenomena remains elusive.

*Ferroelectricity*: In materials, in which the para- to ferroelectric transition is driven by a soft mode, phononic rectification provides a path to switching the electric polarization [14]. In ultrafast experiments, only transient reversal of the polarization has so-far been achieved, due to macroscopic restrictions such as the strong depolarization field. Theory has suggested workarounds including depolarization field screening and multi-pulse excitations to permanently reverse the polarization [15], [16]. Recent measurements show that some of these problems may be avoided when driving phonons at the longitudinal-optical frequency, where the dielectric function becomes zero [17]. In addition to switching, experiments and theory showed that driven phonons can induce a polarization in paraelectric materials, such as $SrTiO_3$ [18], and even in entirely nonpolar materials [19,20]. The primary challenge for the coming years is to predict mechanisms feasible in experiment and to extend them to 2D materials [21].

*Magnetism*: The control of magnetic order through coherently excited phonons has been dubbed "phonomagnetism". Two mechanisms are dominant here: First, modifying magnetic interactions through structural distortions caused by phononic rectification. Theoretical predictions suggest light-induced transitions between different magnetic orders [22,23], but experimental verifications have remained scarce. Second, coupling effective magnetic fields produced by chiral phonons to the electron spin. Here, the situation is inverted: early theoretical predictions have underestimated the effective magnetic fields by orders of magnitude, typically on the order of microtesla [24,25], whereas experiments have turned out to yield militesla- or even tesla-strength fields [34,35]. Promising developments in turn included theory based on orbit-lattice and electron-phonon coupling predicting or verifying the experimentally found magnitudes [8,9]. Future research will have to include a unified theory of phonomagnetic fields as well as a systematic search for materials exhibiting large couplings that allow for light-induced magnetic phase transitions to occur.

*Chirality/ferroaxiality*: An emerging frontier in phononics is the creation and control of unconventional orders. One example is ferroaxial order—a rotational structural order breaking neither parity nor time-reversal symmetry. The absence of symmetry breaking results in a vanishing conjugate field, rendering external control challenging. However, recent theory proposes a nonlinear phononics pathway to overcome this limitation [26]. Similarly, chiral order has garnered interest due to its inherent immutability in solids. This order arises during the liquid-to-solid transition and remains fixed due to its high energy scale, as seen in materials like quartz. Building on the principles of inducing ferroic orders, protocols utilizing phononic rectification have recently been developed to induce chirality in solids [27]. Experimental applications of these protocols appear to validate this concept [28]. These developments open an entirely new direction within the field, promising applications such as ultrafast switches for chirality-dependent phenomena, e.g., the chirality-induced spin selectivity effect.

**Advances in Science and Technology to Meet Challenges**

*Theory development*: Most of the existing literature deals with excitations and nonlinear couplings of phonons at the Brillouin-zone center. This covers only a slice of the nonlinear dynamics generally possible in solids. Future developments will therefore have to include high-momentum phonons stretching the entire reciprocal space, which can be triggered through nonlinear excitation pathways. Recent theory suggested that high-momentum excitations could result in translational-symmetry breaking in solids [29,30]. Furthermore, nonlinear effects can be induced not just by light, but also by strong interactions of phonons with vacuum electric fields in optical cavities. It has been suggested that nonlinear frequency-conversion processes could be manipulated in cavities, leading to enhanced sum-frequency excitation and parametric amplification [31,32]. We expect these results to spark more predictions of phenomena associated with high-momentum and cavity phonon dynamics that might be measured by cutting-edge pump-probe experiments.

*Computational methodology*: While theory development will lead to the prediction of new fundamental physics, computational methodology will have to keep track to translate the theories into numbers and quantitative predictions for real materials. A particularly important role will be played by machine-learned force-field methods that are already revolutionizing the efficiency of molecular dynamics simulations with ab initio accuracy and that will be essential for comprehensive simulations of the nonlinear phonon dynamics. A key challenge is posed by the integration of the dynamics of other coupled collective modes, such as magnons, plasmons, or excitons, on an equal footing to the lattice dynamics. At the same time, increasing computational power in supercomputing centers has made simulations of time-dependent density functional theory feasible for small systems and picosecond timescales relevant for phonon dynamics [18,33]. A combination of these methods will enable simulating coupled dynamics with yet unknown efficiency and accuracy.

*New and old materials*: While methodological developments are important to discover new effects and refine calculations, the field of nonlinear and chiral phononics still displays a big knowledge gap regarding materials diversity. There is no general physical intuition yet for when materials exhibit large nonlinear phononic couplings or phonon magnetic moments. Such intuition would be necessary, however, in order to achieve all-optical control over materials properties by the means of phononic processes. The reason for this knowledge gap: there simply is only a handful of materials that have yet been investigated both experimentally and theoretically. Hence, a systematic large-scale investigation of materials is timely, for example facilitated by high-throughput first-principles calculations, that will help to build physical and chemical intuition and outline design principles for "good" nonlinear and chiral phononic materials.

## Concluding Remarks

The field of nonlinear and chiral phononics is booming, as tabletop terahertz to mid-infrared pump-probe setups are becoming widely available and the discovery of phonon-induced phenomena accelerates. Developments in chiral phononics face additional challenges in generating high-field circularly polarized radiation, which need to be overcome in the next years. Theory plays an important role in creating an understanding for the phenomena seen in experiment, but even more critically in predicting new physical effects and pathways to controlling macroscopic electronic phases that could pave the way for technological applications. For chiral phononics, the journey has just begun, and exploring the impact of phonon angular momentum on the electronic system not only opens a door towards directly accessing spin-, orbit-, and valley-dependent properties, but also revisits our basic understanding of angular momentum coupling in solids. A concerted theoretical, computational, and experimental effort is therefore timely to push this field on and towards phono-electronic and phono-spintronic devices in the future.

## Acknowledgements

This work was supported by Eindhoven University of Technology and MPSD.

## 13 – Towards FAIR time-dependent data from time-resolved experiments

**Laurenz Rettig,** Fritz Haber Institute of the Max Planck Society, Faradayweg 4-6, 14195 Berlin, Germany.
rettig@fhi-berlin.mpg.de

**Ralph Ernstorfer,** Fritz Haber Institute of the Max Planck Society, Faradayweg 4-6, 14195 Berlin, Germany. Institut für Optik und Atomare Physik, Technische Universität Berlin, Straße des 17 Juni 135, 10623 Berlin, Germany.
ernstorfer@tu-berlin.de

**Status**

Data have been termed the "oil of the 21st century", unlocking great potential if refined and processed in the right way. In scientific research, particularly modern data science concepts like artificial intelligence (AI), machine learning, or neural networks enable novel types of data analysis with often strong predictive power [1,2]. While offering great benefits, such methods require large amounts of well-characterized scientific data as input, fueling a growing demand for publicly accessible research data following the FAIR principles (findable, accessible, interoperable, reusable) [3], also referred to as "findable and AI ready" [4]. This is enabled by rich metadata annotation, standardized file formats and powerful metadata search engines, as demonstrated recently by databases for *theoretical* solid-state physics research data like the NOMAD project [5]. In contrast, for *experimental* research such standardization, metadata schema, and public databases are still scarce.

An excellent example of the impact of FAIR data principles is given by structural databases such as the Protein Data Bank (PDB) [6] or the Inorganic Crystal Structure Database (ICSD) [7]. These repositories have transformed their respective fields by providing high-quality, standardized experimental datasets that are easily accessible and reusable. This is a prerequisite for the development of powerful deep learning algorithms like AlphaFold [8], the protein structure prediction tool at the core of the Nobel Prize in Chemistry 2024. Similarly, establishing well-curated, openly accessible datasets for time-resolved spectroscopies can enable future data-driven discoveries. Access to such properly documented experimental datasets enables advanced data analysis techniques, often in combination with theoretical models, to uncover new insights. Such a synergy between experiments and theory can accelerate scientific breakthroughs, and offer a deeper understanding of dynamic processes at the microscopic level.

Ultrafast spectroscopy and imaging present an extraordinarily challenging domain for FAIR data standardization as these techniques probe *nonequilibrium* states of matter, yielding a potentially infinite variety of states for any given material. Additionally, setups for ultrafast time-resolved experiments are predominantly custom-built, with considerable heterogeneity in both hardware and software design, further complicating such efforts.

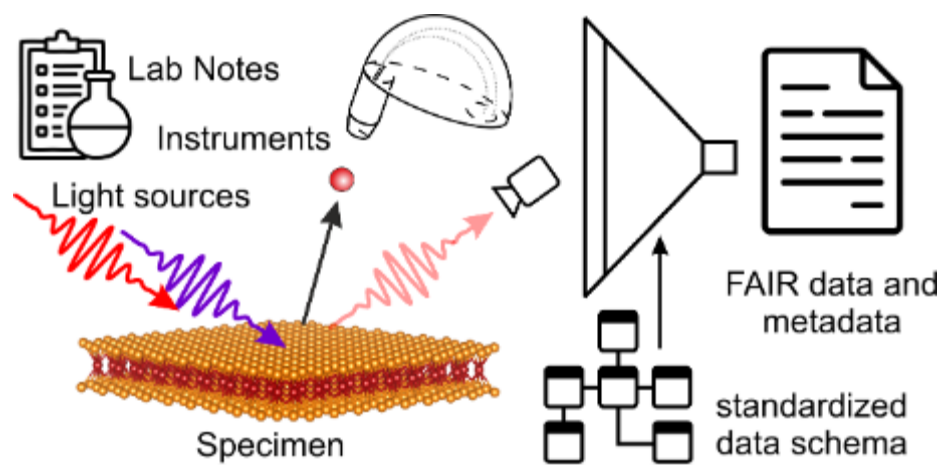

*Figure 1.* Workflow for generation of FAIR experimental data of ultrafast experiments.

**Current and Future Challenges**

One of the biggest challenges in implementing FAIR data principles for experimental data is the "I" in FAIR, i.e. interoperability. The challenge here lies in integrating data from different sources, including human-generated metadata, into a structured and well-documented form (see Fig. 1). In particular time-resolved experiments, which often involve complex setups and multiple measurement techniques and instrumentation, typically employ highly diverse data formats, including custom-designed and vendor-specific data structures. Ensuring that these datasets can be easily integrated and interpreted across different platforms and by various research groups requires a high degree of standardization driven by a certain user community, which is currently often lacking. This issue is exacerbated by unavailable, inconsistent, or unstructured metadata, often due to the use of proprietary software or closed vendor formats that restrict data sharing. Furthermore, essential experimental details, such as laser pulse parameters, sample history, or experimental geometries, are often inadequately captured, especially when lab notes are recorded in traditional, unstructured (paper) lab notebooks, or missing entirely often enough.

A key step towards solving these issues is the harmonization of data schemas, glossaries of terms, or ontologies. A unified framework for describing time-resolved spectroscopic data would greatly improve interoperability by ensuring that researchers (and their instruments) are speaking the same language when it comes to data and metadata. In addition, such FAIR ultrafast data would also strongly benefit theoretical efforts, both as a starting point for model development and as reference benchmark data for model verification. However, achieving this harmonization requires significant community-wide efforts, including the development of research databases that support rich standardized metadata and offer advanced search functionalities. Combined with online analysis platforms, this would enable more precise data retrieval and reuse, and facilitate integration between theory and experimental workflows.

A related challenge is ensuring community acceptance of these standardized practices. While the technical infrastructure for FAIR data is essential, the cultural shift within the scientific community to adopt these practices is equally important. Researchers must be incentivized to document their experiments rigorously, share data openly, and adopt new tools for ensuring data interoperability.

In addition to these broader challenges, time-resolved spectroscopies face specific technical difficulties in creating structured datasets. For example, accurately describing pulse shapes, spot profiles, and fluence is essential for replicating experiments, robust analyses, and establishing benchmarks. However, these details are often insufficiently described or inconsistently recorded. Addressing these gaps will require not only consistent metadata standards but also new tools and interfaces that make it easier to capture complex experimental parameters in a structured, reusable format.

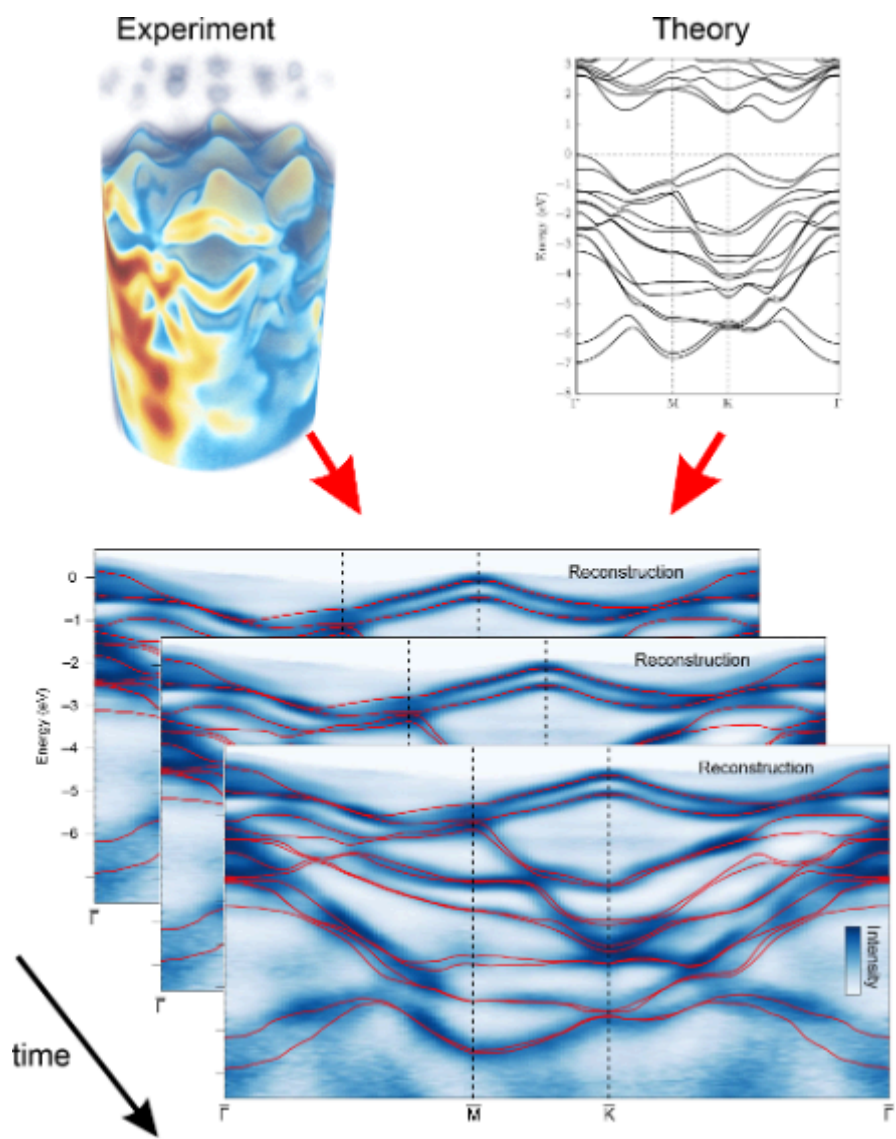

**Figure 2**. Example workflow for theory-informed extraction of time-dependent band structure information from trARPES data. Adopted from [16].

**Advances in Science and Technology to Meet Challenges**

Various developments in science and technology will be required to address the challenges in implementing FAIR data principles for time-resolved spectroscopies. Already now, initiatives like the NFDI (National Research Data Infrastructure) [9], DAPHNE4NFDI [10] or FAIRmat [11] are driving efforts to create standardized frameworks for curating, managing and sharing scientific data. These initiatives aim at developing data schemas and formats, as well as data management platforms that facilitate the structured acquisition and sharing of complex datasets. Recent examples of such efforts include the definition of (meta) data schemas based on the NeXus format [12] for (time-resolved) optical spectroscopy or time- and angle-resolved photoemission spectroscopy (trARPES) with a rich set of well-defined metadata, supported by the data management platform NOMAD [13] (see Section 14).

While such efforts are promising, further developments are needed to meet the demand for a convenient user experience that is the basis of an adoption by the community. Here, one of the most promising advances is the use of AI-based tools for the automated collection and conversion of data and metadata, supported e.g. by recent advances in large language models (LLMs). Such tools could significantly reduce the burden on researchers by automatically extracting experimental parameters and metadata, ensuring that data is consistently documented in a structured, reusable format. This helps bridge the gap between the demand for convenient interfacing and the need for structured data acquisition, making it easier to implement FAIR principles without disrupting experimental workflows.

Finally, the development of advanced, AI-driven search tools offers significant potential for improving data discoverability. These tools could enable researchers to perform more sophisticated queries across large, complex datasets, including context-dependent selection on domain-specific metadata. By leveraging these technological advancements, the adoption of FAIR principles in experimental solid-state research, and time-resolved spectroscopies in particular, can be streamlined, enabling novel research paradigms.

**Concluding Remarks**

While at the current stage, established metadata standards, best-practice workflows, and general adoption of FAIR principles for time-resolved data are still scarce, a strong push towards establishing such routines exists from the community, as well as from funding and publishing agencies. Whereas even from a single lab, FAIR data simplifies comparison of datasets from different experiments, widespread availability of such FAIR time-resolved data will hugely benefit both experimental and theoretical research, and enable new types of analysis based on machine-learning approaches. Recent highlight examples include e.g. the deep-learning based analysis of time-resolved luminescence data [14] or the AI-based interpretation of ultrafast magnetization dynamics [15]. Moreover, new analysis approaches could bridge the gap between experimental and theoretical efforts to bring these disciplines closer together, e.g., the theory-informed, machine-learning based determination of the transient electronic band structure after photoexcitation from trARPES experiments (Fig. 2) [16]. Combined with molecular dynamics simulations [17] or time-dependent dynamical mean field models [18], such analysis could provide unprecedented insight, e.g., into photoinduced phase transitions [17,19], mode-resolved electron-phonon coupling [20], transient band gap renormalization [21], or ultrafast Floquet engineering of band structures [22]. Such advances will not only improve data-driven discoveries, but also foster collaboration and innovation, ultimately pushing the boundaries of ultrafast solid-state research.

**Acknowledgements**

This work received funding from the Max Planck Society. The authors acknowledge funding from the Deutsche Forschungsgemeinschaft (DFG, German Research Foundation) within the FAIRmat consortium, project 460197019.

# 14 – FAIR data management for time-resolved theoretical spectroscopy

**Joseph F. Rudzinski**, Physics Department and CSMB Adlershof, Humboldt-Universität zu Berlin, 12489 Berlin, Germany
joseph.rudzinski@physik.hu-berlin.de
**Ronaldo Rodrigues Pela**, Supercomputing Department, Zuse Institute Berlin (ZIB), Berlin, Takustraße 7, 14195 Berlin, Germany
ronaldo.rodrigues@zib.de
**Claudia Draxl,** Physics Department and CSMB Adlershof, Humboldt-Universität zu Berlin, 12489 Berlin, Germany
claudia.draxl@physik.hu-berlin.de

## Status

Insights from theoretical spectroscopy can significantly enhance our understanding of light-matter interaction and the elementary processes happening on the electronic scale. *Ab initio* methods based on many-body theories are able to describe and accurately predict charged excitations such as photoemission as well as neutral excitations probed by optical absorption, core-level and Raman spectroscopy, and alike. As a consequence, methods beyond density-functional theory (DFT) have gained prominence in theoretical spectroscopy. Since many excitations can be expressed as the response of a system to electromagnetic radiation, these methods are often based on linear-response theory. Prevalent methodologies in this context are the *GW* method and the Bethe-Salpether equation (BSE) of many-body perturbation theory (MBPT) as well as time-dependent DFT (TDDFT). Such methods are already available within several *ab initio* codes.

*GW*, BSE, and TDDFT calculations are much more intricate than DFT. Apart from scaling worse with the number of atoms, they typically converge more slowly with respect to parameters controlling numerical precision, e.g., the vacuum length in low-dimensional systems or the k-point sampling, and require further convergence parameters, such as unoccupied states. Other methodological aspects can also severely impact the accuracy of excited-state calculations: the DFT starting point or the degree of self-consistency in *GW*; the exchange-correlation kernel in TDDFT; the static or dynamic screening and the inclusion or exclusion of resonant-antiresonant coupling in BSE. Consequently, achieving reproducibility and interoperability across different calculations becomes formidable. For example, in the case of monolayer $WSe_2$—a very well-studied, rather simple 2D material—published $G_0W_0$ band gaps scatter in the range of 1.70 to 2.89 eV, i.e., with more than 1.1 eV deviation [1], related to variations in numerical approaches, spin-orbit coupling, DFT starting-point, and underlying geometry. Similar scenarios have also been observed for TDDFT [2]. Recent developments in describing dynamical processes dramatically magnify these complications.

Beyond reproducibility issues, theoretical spectroscopy poses broader challenges in the context of data management. The biggest difficulty is arguably metadata capture, complicated by an array of methodology choices, numerical implementations, and computational parameters. These calculations, associated with complex workflows, also demand extensive computational resources and produce large datasets that are difficult to store in their full complexity. Additionally, multiple public and in-house software for executing these calculations exist, complicating both metadata interconversion and comparison of results between programs. To overcome these challenges, immediate adoption of the FAIR (Findable, Accessible, Interoperable, and Reusable) data principles [3] is required, enabling essential knowledge extraction from raw computational data, enhancing verifiability and reproducibility, and supporting data-driven approaches for their reuse. This approach holds the potential to dramatically accelerate discoveries and innovations across various domains [4].

In general, public databases within the materials science community, for example Refs. [5,6], represent a meaningful step towards findability and accessibility considerations. The open-source web service NOMAD [7], a pioneer in the application of the FAIR data principles to the organization, analysis, sharing, and publishing of *ab initio* data, has emerged as a customizable data infrastructure, providing tools for diverse computational and experimental techniques, as well as domain-specific use cases. NOMAD already supports several theoretical-spectroscopy methods, including *GW* and BSE.

**Current and Future Challenges**

Despite the success of *GW*, BSE, and TDDFT, assuming a linear response to model light-matter interaction can be a severe approximation, especially in experiments with ultrafast laser pulses, which have become widespread in the past years. Examples of such pump-probe experiments are time-resolved optical spectroscopy, time-resolved resonant inelastic x-ray scattering (RIXS) [8,9], and time- and angle-resolved photoemission spectroscopy [10]. They allow for studying electron excitation dynamics, charge and exciton propagation, as well as electron-phonon and exciton-phonon couplings [11]. In such scenarios, the nonequilibrium regime prevails, requiring insights from theory beyond the (static) linear-response picture. One solution is to use TDDFT in the time domain (real-time TDDFT, RT-TDDFT), for example to study electron dynamics under the influence of strong laser pulses. RT-TDDFT can also be coupled with the nuclear motion, either in a semi-classical picture or in a full quantum mechanical approach [12]. An alternative is to resort to nonequilibrium Green's functions, whose real-time evolution follows the Kadanoff-Baym equations. This gives rise to approaches like time-dependent BSE [13], time-dependent adiabatic *GW* [14], *GW* plus real-time BSE coupled with molecular dynamics (MD) [15], and the real-time *GW*-Ehrenfest-Fan-Migdal method [16]. Other recent advancements include theoretical methods designed to account for electron-phonon [17] and exciton-phonon couplings, enabling the study of geometry relaxation in excited states [18] and, thus, a path for treating photo-luminescence.

Each of these approaches tremendously complicates (meta)data capture, introducing not only an additional variable, time, but also more intricate processes and, hence, quantum-mechanical operators. Moreover, the combination of distinct methodologies into various complex methods requires a hierarchical workflow description with complex metadata for every individual task and elaborate connections between them.

**Advances in Science and Technology to Meet Challenges**

Over the past two years, NOMAD has transformed from a domain-specific repository into a global platform for customized FAIR data management in materials science and related fields. NOMAD addresses simulation interoperability challenges through a comprehensive core simulation schema [19] that encompasses a wide range of techniques, and provides semantic context by linking to a materials property taxonomy [20]. The backbone of NOMAD's compliance to the FAIR principles is its structured data schema: the NOMAD *Metainfo*. The Metainfo provides a hierarchical and modular structure for metadata (Reusability), with unique naming and human-readable descriptions (Findability), along with categorical typing for linking to ontologies (Interoperability) and persistent identifiers that can be accessed programmatically (Accessibility) [21]. Designed for extension and customization, NOMAD's plugin-based system allows users to create parsers (i.e., self-contained Python modules for extracting (meta)data from simulation files) and schemas for their specific needs. NOMAD also offers a generalized data schema for workflow graphs, ensuring accurate provenance documentation, and facilitating the curation of AI-ready datasets. These advances in data infrastructure are strongly in line with proposed roadmaps for accelerating materials discovery through machine learning [22] and meeting future challenges for exascale computing [23].

The development of infrastructure and tools alone cannot overcome interoperability and reproducibility challenges in theoretical spectroscopy. The community must adopt common data structures, like schemas, and higher-level semantic characterizations, such as taxonomies or ontologies, while promoting knowledge exchange and data management education, especially for younger scientists. The consortium FAIRmat [24]—developer of the NOMAD infrastructure—supports these efforts through tutorials, workshops, and hack-a-thons. For example, a CECAM flagship workshop on "FAIR and TRUE Data Processing for Soft Matter Simulations" in September 2023 brought together diverse members of the MD community, including FAIR-data consortia, simulation engine developers, institutional data stewards, HPC resource managers, and researchers focused on FAIR data standards. To build upon these efforts, a hack-a-thon for "FAIR Data Management of Theoretical Spectroscopy and Green's Function Methods" was recently held, providing hands-on guidance in developing parsers and schemas using the NOMAD software. Continued outreach in this direction is essential for advancing data management in this field.

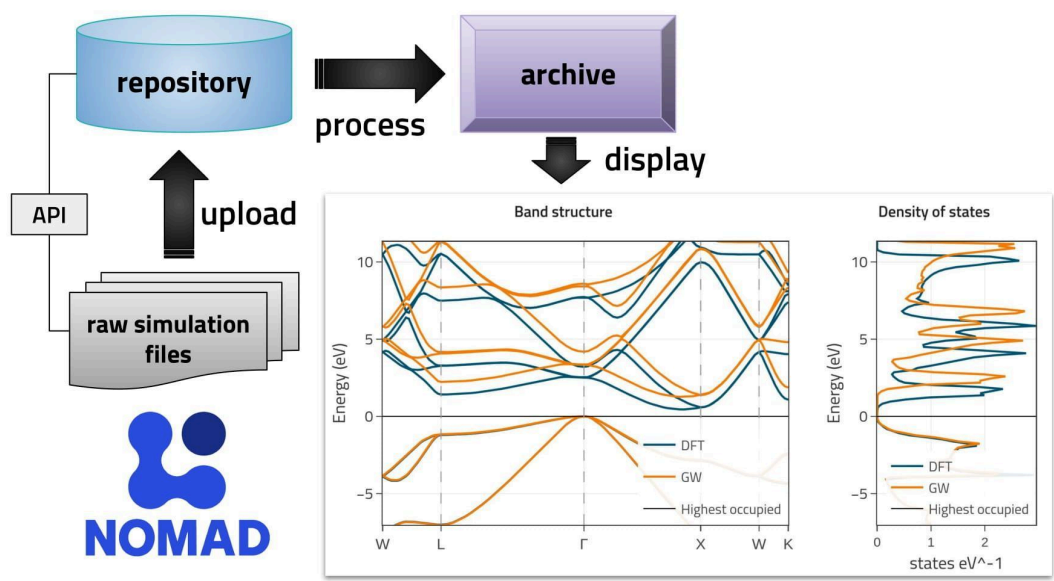

**Figure 1.** Schematic of NOMAD's processing and visualization pipeline, along with a screenshot of the results from a *GW* calculation, displayed within NOMAD's GUI. NOMAD automatically recognizes raw simulation files that contain *GW* calculations from supported simulation codes, locates the DFT starting point used, extracts all relevant metadata from both sets of files, stores them within a structured schema that is agnostic to the particular code used, and plots scientifically relevant outputs (e.g., comparisons of the DFT and *GW* electronic structure, as shown) for immediate user inquiry.

**Concluding Remarks**

Time-resolved theoretical spectroscopy is not only challenging for the advancement of theory but also for proper data handling. These challenges arise due to the inherent complexity of the processes to be described, as well as the high level of accuracy and enormous computational resources required. In order to guarantee that the data is FAIR, the metadata—the details of methods and approximations, numerical implementations and computational parameters, starting points underlying high-level calculations, etc.—need to be meticulously documented. Data infrastructures should also provide insight through visualization and data analysis, thus going well beyond what can be published in research journals. In combination with FAIR experimental data (see section 13), these data management approaches should also provide a framework for model development and enable rigorous benchmarking of theories.

NOMAD provides a powerful framework for addressing these challenges. It supports MD simulations with storage and visualization of trajectory, time-dependent and ensemble-averaged properties, as well as system hierarchies and molecular topologies. NOMAD has over 14,500 MD entries, each containing a full simulation trajectory. NOMAD also supports advanced MBPT calculations, including *GW*, BSE, and dynamical mean-field theory (DMFT). It not only automatically recognizes and stores metadata for "standardized" theoretical spectroscopy workflows (see Fig. 1), including DFT+*GW*, DFT+*GW*+BSE and DFT+TB+DMFT, but also allows more complex, custom workflows via user-defined schemas. Users can connect individual simulations or tasks together, enabling extendable and easily specialized documentation of complex simulation techniques. NOMAD hosts over 23,000 entries containing *GW*, BSE, and DMFT data, along with the provenance of their workflows. Additionally, an increasing number of detailed methodological filters, e.g., *ab initio* precision filtering, are being implemented into NOMAD, increasing the specificity of possible data queries. Finally, NOMAD offers a unified approach for FAIR handling of both experimental and computational spectroscopy data. Thus, NOMAD demonstrates immense potential as a holistic data management solution for tackling current and future challenges within the theoretical spectroscopy community.

## Acknowledgements

This work is funded by the NFDI consortium FAIRmat - Deutsche Forschungsgemeinschaft (DFG) - Project 460197019.